\documentclass[12pt, english]{article}
\usepackage{comment}
\usepackage{natbib}
\usepackage{algorithm,algorithmic}%
\usepackage[leqno]{amsmath}
\usepackage{mathrsfs}
\usepackage{amssymb}
\usepackage{mathtools}
\usepackage{bm}
\usepackage{ctable}
\usepackage{setspace}
\usepackage{longtable}
\usepackage{url}
\usepackage{moredefs,lips} 
\usepackage{IEEEtrantools}
\usepackage{multirow, bigdelim}
\usepackage{enumerate}
\usepackage[normalsize, bf]{caption}
\usepackage{afterpage}
\usepackage[all]{nowidow}
\usepackage{listings}
\usepackage[margin=1in]{geometry}
\usepackage{subcaption}
\usepackage{graphicx}
\usepackage[title]{appendix}
\title{A Hybrid Approach to Targeting Social Assistance\footnote{
We would like to thank Michael Hillebrecht and Stefan Klonner for 
providing the data from Burkina Faso and graciously answering our
questions about the dataset.
We further appreciate the useful feedback provided by participants of
the 2021 Annual Meetings of the Southern Economic Association and 
the 2021 Innovations for Poverty Action and Global Poverty Research Lab 
Methods and Measurement Conference.}}

\author{
Lendie Follett\footnote{
Email: lendie.follett@drake.edu.}
~\\
Department of Information Management and Business Analytics \\
Drake University \\
~\\
Heath Henderson\footnote{
Corresponding author;
325 Aliber Hall, Des Moines, IA 50311;
Tel: +1 515 271 2898; Email: heath.henderson@drake.edu.}  
~\\
Department of Economics and Finance \\
Drake University}

\date{\today}

\begin{document}


\begin{titlepage}
\maketitle
\begin{abstract}
Proxy means testing (PMT) and community-based targeting (CBT) are two of the 
leading methods for targeting social assistance in developing countries.
In this paper, we present a hybrid targeting method that incorporates CBT's 
emphasis on local information and preferences with PMT's reliance on verifiable 
indicators.
Specifically, we outline a Bayesian framework for targeting that resembles PMT 
in that beneficiary selection is based on a weighted sum of sociodemographic 
characteristics.
We nevertheless propose calibrating the weights to preference rankings from community 
targeting exercises, implying that the weights used by our method reflect how 
potential beneficiaries themselves substitute sociodemographic features when making 
targeting decisions.
We discuss several practical extensions to the model, including a generalization 
to multiple rankings per community, an adjustment for elite capture, a method for 
incorporating auxiliary information on potential beneficiaries, and a dynamic 
updating procedure.
We further provide an empirical illustration using data from Burkina Faso and 
Indonesia. \newline
~ \newline
\noindent
\emph{Key words}: Bayesian inference; community-based targeting; 
proxy means testing; social assistance; targeting  \newline
\textit{JEL codes}: C11; D04; I32; I38; O20
\end{abstract}
\end{titlepage}
\newpage
\onehalfspacing


\section{Introduction}
\label{sec: intro}

Effectively identifying the poor is critical to the success of many social
assistance programs, particularly in developing countries.
Proxy means testing (PMT) is a popular method for ``targeting'' social 
assistance and has been implemented in a wide variety of countries
\citep{coady2004, fiszbein2009, devereux2017}.
The standard form of PMT ranks potential beneficiaries on the basis of a weighted 
sum of sociodemographic characteristics, where the weights are coefficients from a 
regression model of income or expenditure estimated using household survey data 
\citep{kidd2011, kidd2017, brown2018}.
PMT is easy to implement, relies only on verifiable indicators, and imposes limited 
costs on potential beneficiaries.

Though PMT has a number of strengths, it also has limitations.
As a highly centralized form of targeting, PMT is inconsistent with commitments to 
participatory development.
In particular, PMT conflicts with Target 16.7 of the Sustainable Development Goals, 
which aims to ``[e]nsure responsive, inclusive, participatory and representative
decision-making at all levels'' \citep{unga2015}.
PMT also neglects local information, as household surveys may have 
limited geographic coverage of program areas, disproportionate non-response 
from the poor, and may not capture all relevant individual or household 
features \citep{alderman2002local, carr2013missing, 
bollinger2019trouble}.
A further concern is that PMT may conflict with local preferences regarding 
beneficiary selection.
Perhaps most importantly, the income-oriented definition of poverty used by 
PMT may be inconsistent with local definitions of poverty 
\citep{alatas2012targeting, han2019community, hillebrecht2020community}.%
\footnote{See \citet{kidd2017} or \citet{brown2018} for discussion of additional
limitations of PMT.}

Community-based targeting (CBT) is a leading alternative to PMT.
CBT typically revolves around an exercise where community members or local 
leaders meet to rank potential beneficiaries in terms of need
\citep{conning2002community}.
CBT is a highly participatory targeting method that emphasizes local 
information and preferences, but it is also subject to limitations.
While CBT is often considered less costly than PMT, the cost savings
are achieved by reducing reliance on remunerated labor (e.g., for data 
collection and processing) and imposing additional unremunerated costs on community 
members in the form of opportunity and psychological costs \citep{devereux2017}.%
\footnote{The opportunity costs associated with CBT result from the fact that 
local officials and community members must engage in extended exercises
for identifying needy households and thus forego devoting time to productive 
activities.
In Burkina Faso, for example, \citet{hillebrecht2020community} found that the
CBT exercises took a half a day for each community on average.
The psychological costs of CBT are less acknowledged, but no less important.
In particular, community members may find it shameful or embarrassing to be
publicly identified as poor during the community ranking exercises.
They may also find it difficult to discuss private matters (e.g., their 
standard of living or the standard of living of others) in a public
setting.}
In addition to imposing costs on community members, it is well-known that 
CBT can suffer from elite capture, where local elites use their power to 
influence the beneficiary selection process \citep{bardhan2006pro,
kilic2015decentralised, han2019community, basurto2020decentralization}.

In this paper, we develop a hybrid targeting method that incorporates many of
the advantages of PMT and CBT while minimizing their main limitations.
Specifically, we outline a Bayesian framework for targeting that resembles
PMT in that beneficiary selection is based on a weighted sum of 
sociodemographic characteristics, but we instead propose calibrating the weights to 
community preference rankings.
The ratio of any two weights then reflects the implied rate at which potential 
beneficiaries themselves substitute sociodemographic features when making 
targeting decisions.
Our approach thus inherits the advantages of PMT (e.g., beneficiary selection 
is based on objective and verifiable information), but discards its less 
desirable aspects.
That is, in contrast to PMT, our method broadly respects commitments to participatory
development, privileges local information, and is consistent with preferences
revealed through community ranking exercises.

Our hybrid approach also overcomes the main limitations of CBT.
While CBT is costly to potential beneficiaries in the sense that many (if not all)
community members must participate in community ranking exercises, our method
can be calibrated with community ranking data from a small sample of communities. 
Program administrators can thus minimize the opportunity and psychological costs
imposed on beneficiaries by conducting model-based prediction of community preference
rankings in non-sampled locations.
Regarding elite capture, an important feature of our approach is that it 
explicitly models the influence of elite connections on community rankings and
provides a simple way to purge these influences from the final ranks.
The procedure thus mitigates concerns about transmitting the influence
of elite capture to non-sampled communities and can also be used to reconstruct 
unbiased rankings from sampled communities.

There are other features of our approach worth highlighting.
First, our model provides a principled way to aggregate rankings when CBT exercises 
generate multiple rankings per community.
In Niger, for example, each community constituted three committees to independently
rank potential beneficiaries, but then the ranks were aggregated in an \emph{ad hoc}
manner to arrive at the final list of beneficiaries \citep{premand2020efficiency}.%
\footnote{Specifically, the final list of beneficiaries was based on each household's
average rank from the three committees.
See \citet{hillebrecht2020community} for another example of CBT with multiple 
rankings per community.}
Second, as a fully Bayesian framework, we strategically use priors to regularize 
the model's coefficients to improve out-of-sample predictions, to incorporate
subjective beliefs on the quality of different rankers, and to provide an
efficient way to update the model over time.
Finally, our framework provides a way to introduce auxiliary data (e.g., information
on incomes or expenditures) to improve performance when preference rankings are only 
generated for a small number of communities.

We illustrate our method using data from Burkina Faso and Indonesia. 
These datasets were previously used by \citet{hillebrecht2020community} (Burkina Faso)
and \citet{alatas2012targeting} (Indonesia) to compare the performance of alternative 
targeting methods, and therefore combine information from community ranking exercises
with household survey data.
We use the datasets to (1) examine how our method weights sociodemographic
characteristics relative to PMT, (2) demonstrate the out-of-sample predictive 
performance of the method, and (3) illustrate the various model features mentioned 
above.
We find that communities implicitly weight sociodemographic characteristics
quite differently than PMT, often with practically-relevant differences in 
signs and magnitudes.
We further find that our method performs well in terms of predicting community
preference rankings, most notably achieving error rates that are lower than what 
PMT can achieve when predicting household expenditures.

This paper contributes to the methodological literature on targeting social assistance.
A few recent studies have suggested using PMT with alternative estimands, including 
food expenditures \citep{basurto2020decentralization}, dietary diversity
\citep{premand2020efficiency}, and human capabilities \citep{henderson2022}.
Other recent work has proposed using alternative estimators to improve the
predictive performance of PMT \citep{mcbride2018retooling, brown2018}.
Finally, \citet{alatas2012targeting} and \citet{stoeffler2016reaching} examined the
performance of a hybrid targeting approach where the selected beneficiaries were 
those with the lowest PMT scores among a larger group of households selected via CBT.%
\footnote{Also see \citet{barrett2003accurate}, \citet{elbers2007poverty}, and 
\citet{coady2009targeting} for some discussion of the potential complementarities 
between different targeting approaches.}
For example, \citet{alatas2012targeting} had communities nominate 1.5 times 
the quota of beneficiaries and then narrowed the list using PMT.

Our hybrid approach is distinct from existing hybrid methods. 
Rather than using CBT rankings to narrow the pool of potential beneficiaries for 
which PMT is applied, we instead (1) estimate the model's weights using 
community preference rankings and (2) use information on incomes or expenditures to
improve the predictive accuracy of the model.
That is, the distinguishing feature of our approach is that we use 
community revealed preferences as the estimand, whereas existing hybrid approaches
focus on incomes or expenditures.
While emphasizing community preferences aligns our method with commitments to 
participatory development, it is also better reflects local information and values 
(e.g., local definitions of poverty), which may improve community satisfaction with 
targeting outcomes and reduce the social unrest often associated with 
statistical targeting procedures.%
\footnote{A number of papers find that local definitions of poverty are often
distinct from standard income-oriented definitions of poverty.
For examples of such findings, see \citet{alatas2012targeting} on earning
capacity, \citet{han2019community} on multidimensional poverty, and 
\citet{hillebrecht2020community} on assets (among other dimensions).
\citet{alatas2012targeting} find that targeting procedures that are in line
with community preferences (i.e., CBT) tend to improve community satisfaction, 
though \citet{premand2020efficiency} find that satisfaction can be undermined
if community members believe that CBT exercises are manipulated for personal
gain.
Finally, it is well-documented that PMT is often associated with social unrest 
due to lacking transparency, inaccuracy, or the fact that it conflicts with local
definitions of poverty \citep{cameron2014, kidd2017, sumarto2021welfare}.}

In what follows, Section \ref{sec: methods} describes a basic version of our 
Bayesian framework for hybrid targeting and Section \ref{sec: extensions} 
presents various model extensions.
Section \ref{sec: data} then discusses the data we use for our empirical 
illustration and Section \ref{sec: results} details the results of our analysis.
Finally, Section \ref{sec: conclusions} provides concluding remarks, including
a discussion of the limitations of our framework.


\section{Targeting Community Revealed Preferences}
\label{sec: methods}

In this section, we first discuss the data requirements for our method and 
present a simple benchmark model in the probit regression.
Next, we outline the most basic version of our Bayesian framework for 
targeting community revealed preferences, which assumes a singular ranking
scheme for each community and no auxiliary information.
Our focus in this section is on fixing intuition related to modeling
ranked data and, in the next section, we will extend the model in a variety of
ways.

\subsection{Data requirements and benchmark model} 
\label{ssec: probit}

Let $G = \{1, 2, 3, \ldots \}$ represent the set of all geographic units within 
a given country (e.g., neighborhoods, villages, districts, etc.).
We will refer to these units as communities.
As social assistance programs do not necessarily cover all communities within a 
country, we let $J \subseteq G$ denote the set of communities in program areas.
We will index these communities by $j$ and index households within any community 
by $i$.
We then assume that for all households in the program areas we observe a row 
vector of sociodemographic characteristics $x_{ij}$, which is obtained from a 
census of the program areas.
The covariate vector $x_{ij}$ is common to all forms of PMT and is used for 
calculating the scores that determine program inclusion.

Now let $K \subseteq G$ represent the set of communities conducting community 
ranking exercises and index these communities by $k$.
Note that CBT requires $K = J$ or that all program areas perform the ranking 
exercises.
As mentioned above, however, our model only requires that the community ranking
exercises be conducted in a sample of communities, which will most plausibly
be taken from program areas (i.e., $K \subset J$).
In specifying $K \subseteq G$, we nevertheless do not rule out the possibility of 
sampling from non-program areas.
We assume that each exercise ranks households in ascending order from most to
least needy and denote each household's rank by $z_{ik}$.%
\footnote{Note that here we assume that each community only generates a single
list of ranks. 
Below we consider extensions that accommodate multiple rankings per community.}
Finally, we assume that we observe the (same) vector of sociodemographic 
characteristics for all households in the sampled areas, which we denote by 
$x_{ik}$.

At a minimum, we then observe $x_{ij}$, $x_{ik}$, and $z_{ik}$.
Recall that our objective is to develop a PMT-like method where the estimand is
community revealed preferences.
PMT typically consists of two stages: (1) an estimation stage that calibrates
a statistical model and (2) a prediction stage that calculates the scores that
determine program inclusion based on the statistical model.
For example, the first stage in the standard form of PMT uses household survey data 
to regress income or expenditure per capita on some observed sociodemographic 
traits. 
The estimated coefficients are then used in a second stage to predict income or 
expenditure per capita for potential beneficiaries, and it is these scores
that determine program inclusion.
While we retain this simple two-stage procedure, we deviate from the standard
form of PMT by proposing a different first stage where the weights are calibrated 
to community preference rankings.

To this end, first consider a simple benchmark model in the form of a probit 
regression.
Let $d_{ik} = \text{I}(z_{ik} \leq q_k)$ where $\text{I}(\cdot)$ is the indicator 
function and $q_k$ represents the beneficiary quota for community $k$.
The variable $d_{ik}$ is thus a binary variable that captures program inclusion,
which we can model in terms of a continuous latent variable $\tilde{d}_{ik}$ as 
follows:
\begin{equation}
\tilde{d}_{ik} = x_{ik} \beta + \epsilon_{ik}
\end{equation}
where $d_{ik} = \text{I}(\tilde{d}_{ik} > 0)$, $\beta$ denotes a column vector 
of first-stage coefficients (including an intercept), and $\epsilon_{ik} \sim
\text{N}(0, 1)$.
The above is often estimated via maximum likelihood using the following likelihood
function:
\begin{equation}
P(d \mid \beta) = \prod_{k} \prod_{i} (\pi_{ik})^{d_{ik}}
(1-\pi_{ik})^{1-d_{ik}}
\end{equation}
where $\pi_{ik} \equiv P(d_{ik} = 1 \mid x_{ik}) = P(\tilde{d}_{ik} > 0 \mid x_{ik}) 
= \Phi(x_{ik} \beta)$ and where $\Phi(\cdot)$ is the standard normal cumulative 
distribution function.
The likelihood function can then be maximized with respect to $\beta$ and the 
estimated coefficients $\hat{\beta}$ can be used to calculate out-of-sample 
score estimates for all potential beneficiaries as $x_{ij}\hat{\beta}$.%
\footnote{Note that prediction with the probit regression often aims to
estimate the probability that a given observation takes on the value of one.
In our case, this prediction would be calculated as $\Phi(x_{ij}\hat{\beta})$.
Given that $\Phi(\cdot)$ is a monotonic function and that we are only interested 
in relative ranks, we can ignore this additional step.}

While simple, the probit model has important limitations for estimating community 
revealed preferences.
First, the probit model requires dichotomizing the community rankings and thus 
discards information: Beneficiaries and non-beneficiaries near the threshold are 
treated very differently, while all households within the two groups are treated
identically.
Second, the probit model is subject to the issue of complete or quasi-complete 
separation (i.e, when covariates provide perfect or near-perfect predictions), 
which typically requires \emph{ad hoc} solutions to obtain finite maximum likelihood 
estimates.%
\footnote{For example, one common solution is to simply drop the variable
in question, which is often unsatisfactory given the variable's predictive
power.
See \citet{zorn2005solution} for further discussion.}
Finally, the maximum likelihood estimation procedure implicitly assumes a uniform 
prior on all parameters, thus neglecting potentially relevant information that may 
improve out-of-sample predictions and therefore targeting performance.

\subsection{A Bayesian framework}
\label{ssec: basic_model}

Here we outline a basic framework for modeling community rankings that overcomes 
many of the limitations of the probit regression.
The \citet{thurstone1927law} order statistics model is among the most popular for 
modeling ranked data.
The Thurstone model relies on a latent variable formulation similar to the probit
regression, though the latent variable in the Thurstone model determines the ranks 
rather than a dichotomized rank variable.
That is, the ranks are modeled directly, meaning that the Thurstone model not only
makes full use of the available information, but is also not subject to the 
problem of separation.
We specifically consider a class of the Thurstone model called the 
Thurstone-Mosteller-Daniels (TMD) model, which also like the probit assumes that 
the noise associated with the latent variable follows a normal distribution
\citep{thurstone1927law, mosteller1951remarks, daniels1950rank}.

Let $\tilde{z}_{ik}$ denote a continuously-valued latent variable that determines
the observed ranks $z_{ik}$.
We will further use $\tilde{z}_k$ and $z_k$ to denote the vectors of latent variables
and observed rankings for each community, respectively.
We then model $\tilde{z}_{ik}$ as follows:
\begin{equation}
\tilde{z}_{ik} = x_{ik} \delta + \eta_{ik} 
\label{eq:simple}
\end{equation}
where $\text{rank}(\tilde{z}_k) = z_k$, $\delta$ is a column vector of 
coefficients, and $\eta_{ik} \sim \text{N}(0, 1)$.
Note that in the case of ranked data, the parameter vector $\delta$ does not 
include an intercept, as ranks are invariant to the location shifts generated 
by intercepts.
Similar to the probit model, our objective is then to estimate $\delta$, which we 
can then use to calculate the scores for all potential beneficiaries as 
$x_{ij}\hat{\delta}$.

The likelihood is defined as the probability of observing the ranks generated by 
the communities. 
For community $k$, this probability can be represented by $P(z_k \mid \delta)$.
To calculate the likelihood, we must take the conditional likelihood $P(z_k \mid
\tilde{z}_k)$ weighted by the density $P(\tilde{z}_k \mid \delta)$ and integrate 
out the latent variables.
Since the event $z_k$ occurs if and only if $\text{rank}(\tilde{z}_k) = z_k$,
we can write $P(z_k \mid \tilde{z}_k) = \text{I}\big[\text{rank}(\tilde{z}_k) 
= z_{k}\big]$.
We then have the following expression for the likelihood function:
\begin{align}
\nonumber P( z \mid \delta)&= \prod_k P(z_k \mid \delta) \\
\nonumber &= \prod_k \int P(z_k \mid \tilde{z}_k) 
P(\tilde{z}_k \mid \delta) \; d \tilde{z}_k \\
&= \prod_k \int \Bigg \{  \text{I}\big[\text{rank}(\tilde{z}_k) = z_{k} \big]
\prod_i \phi(\tilde{z}_{ik} - x_{ik}\delta) \Bigg \} \;  d\tilde{z}_k
\end{align}
where $\phi(\cdot)$ is the standard normal density function and each integral's 
dimension is equal to the number of ranked individuals in each community.

Maximum likelihood estimation of the above model is non-trivial because of the 
form of the likelihood. 
In particular, evaluation of the likelihood requires numerical approximation
of the integral, which can only be done accurately when the number of ranked
entities is quite small (generally less than 15) \citep{alvo2014statistical}.
To avoid these limitations, we require an alternative estimation procedure
that circumvents the integration.
Bayesian methods have become increasingly popular for estimating Thurstonian
models because they provide a straightforward way to obtain simulation-based
estimates of the unknown parameters \citep{yao1999bayesian, 
johnson2013bayesian}.
Recently, \cite{li2016bayesian} developed a unified Bayesian framework for
estimating a wide array of Thurston-type models and we draw on this work
extensively in what follows.

While maximum likelihood estimation seeks to find the parameter values that
maximize the likelihood of the data, Bayesian inference instead examines the 
probability distribution of the parameters conditional on the data 
(known as the posterior).
The posterior can be derived as follows:
\begin{equation}
P(\delta \mid z) =\frac{P(z \cap \delta)}{P(z)} 
= \frac{P(z \mid \delta) \; P(\delta)}{P(z)} 
\propto P(z \mid \delta) \; P(\delta)
\end{equation}
where $\delta$ represents the unknown parameters and $z$ represents the data.
The first equality above is an application of the conditional probability rule 
and the second equality is Bayes' rule. 
Bayes' rule states that the posterior is proportional to the likelihood 
multiplied by the prior distribution of $\delta$, where the prior describes
our beliefs about the location of $\delta$ before observing the data.
We can thus think of the posterior as a compromise between the information 
contained in the likelihood and the prior (with more weight given to the 
likelihood as more data is obtained).

Bayes’ rule formalizes the process through which we update our beliefs about 
$\delta$ after observing the data.
Priors are central to this process and when we decide on the form of $P(\delta)$
we are balancing the uncertainty of the parameters against the bias.
One could choose ``non-informative" priors that will provide (unbiased) results 
identical to frequentist estimation methods, such as maximum likelihood.
Alternatively, one can place weakly informative priors on $\delta$ based on 
information available \emph{a priori} (further discussed below).
By the bias-variance trade-off, this will reduce the error variability on the 
parameter estimates at the expense of increasing bias.
A critical feature of weakly informative priors in the context of targeting is 
that reducing parameter variability can improve out-of-sample prediction by 
mitigating overfitting \citep{gelman2013, gelman2017prior}.

The Bayesian approach to Thurstonian models circumvents integration through a 
procedure called data augmentation, which facilitates computation
by treating the latent variables as quantities to be estimated
\citep{tanner1987calculation, gelfand1990sampling}.
The augmented posterior can be written as follows:
\begin{align}
    P(\delta, \tilde{z} \mid z) & \propto P(z \mid \tilde{z}) \; P(\tilde{z} 
    \mid \delta) \; P(\delta)  \\ \nonumber 
    &= \Bigg \{ \prod_k \text{I}\big[\text{rank}(\tilde{z}_k) = z_{k} \big] 
    \Bigg \} 
    \Bigg \{ \prod_k \prod_i \phi(\tilde{z}_{ik} - x_{ik}\delta) \Bigg \}
    \text{N}(\mu_\delta, \Sigma_\delta).
\end{align}
where $P(z \mid \tilde{z})$ is the likelihood, the distribution 
$P(\tilde{z} \mid \delta)$ follows directly from our assumptions on $\eta$, and we 
use the multivariate normal distribution for the prior on $\delta$.
In our application of this basic model, we set $\mu_\delta=0$ and 
$\Sigma_\delta = 2.5^2 \times I$ where $I$ denotes an identity matrix.
The prior on $\delta$ reflects the simple belief that most of the covariates are 
unlikely to be useful, but that a few may have a strong relationship 
with community revealed preferences.
The prior thus regularizes the coefficients to avoid overfitting without
resorting to complicated procedures like cross-validation.

The objective is then to draw samples from the posterior and use the samples
to calculate quantities of interest.
We use Markov Chain Monte Carlo (MCMC) methods, specifically the well-known 
Gibbs sampler, to draw samples from the posterior \citep{yao1999bayesian,
johnson2013bayesian, li2016bayesian}. 
Details of the sampling procedure can be found in Appendix \ref{appendixa}.
After obtaining samples, we use the posterior mean of $\delta$ as the weights 
assigned to the sociodemographic characteristics.
That is, letting $b = 1, 2, \ldots, B$ index draws from the posterior, we 
calculate $\hat{\delta} = \frac{1}{B}\sum_{b=1}^B \delta^b$.
As mentioned, the final step is to generate the scores for each potential 
beneficiary as $x_{ij} \hat{\delta}$, and these scores can be used to determine
program inclusion through a rank-ordering of the potential beneficiaries.

It is useful at this point to note how our approach is distinct from 
PMT.
Recall that the standard form of PMT calibrates the weights by using household 
survey data to regress income or expenditures on the observed sociodemographic 
characteristics \citep{brown2018}.
The ratio of any two weights in this case reflects the rate at which the 
corresponding sociodemographic features can be substituted while maintaining a 
given level of income or expenditures.
In contrast, we use community preference rankings as the estimand, meaning that the 
weight ratios associated with our method capture the rate at which community 
members themselves substitute sociodemographic features while maintaining a 
given household rank.
That is, rather than imposing an income-oriented definition of poverty on the 
targeting algorithm, our approach attempts to reflect community preference 
orderings regarding beneficiary selection and thus accommodates alternative 
definitions of poverty.


\section{Extending the Bayesian Framework}
\label{sec: extensions}

In this section, we introduce a number of extensions to our framework for 
targeting community revealed preferences. 
We first extend the framework to accommodate multiple ranking schemes per
community. 
We then discuss the issue of elite capture and present a simple way to remove
the influence of elite capture from the final ranks.
Next, we consider how auxiliary information, namely household survey data,
can be used to improve the predictive performance of the model.
Finally, we outline a procedure for dynamically updating the model's parameters
as new information becomes available.

\subsection{Multiple rankings}
\label{ssec: multiple_rankings}

Some recent implementations of CBT have generated multiple ranking schemes 
per community.
As mentioned, \citet{premand2020efficiency} document a CBT exercise conducted in 
Niger that had each community form three committees: one committee of local leaders, 
one committee of non-leader women, and one committee with a mixed group of
non-leaders.
Each committee conducted its ranking of potential beneficiaries independently 
and then the final ranks for each community were calculated by averaging each
household's rankings across the three lists.
To cite another example, \citet{hillebrecht2020community} discuss a CBT
exercise conducted in Burkina Faso where each community nominated three
key informants to rank potential beneficiaries.
The actual beneficiaries were then selected via an algorithm that prioritized
households based on agreements across the three lists.%
\footnote{That is, any household that was ranked among the poorest by all three
informants was selected as a beneficiary, then households ranked among the poorest
by two informants, and so on.
See \citet{hillebrecht2020community} for detailed discussion.}

To extend our model to accommodate multiple rankers, let $z_{ik}^r$ denote the 
rank of household $i$ in community $k$ by ranking entity $r$, and let 
$\tilde{z}_{ik}^r$ denote the corresponding latent variable.
Further, let $\tilde{z}_k^r$ and $z_k^r$ represent ranker-specific vectors of 
latent variables and observed rankings for each community, respectively.
Similar to \citet{li2016bayesian}, we can then rewrite Eq.\ (\ref{eq:simple}) to 
accommodate multiple rankers as follows:
\begin{equation}
\tilde{z}_{ik}^r = \alpha_{ik} + x_{ik}\delta + \eta_{ik}^r
\label{eq:mrankers}
\end{equation}
where now $\text{rank}(\tilde{z}_k^r) = z_k^r$, $\alpha_{ik}$ is a household-specific
intercept, and $\eta_{ik}^r \sim \text{N}(0, \omega_r^{-1})$.
The term $\omega_r$ represents the precision of the noise in the model, which
is allowed to vary across rankers. 
That is, the model implicitly aggregates alternative ranking schemes according to 
their relative precision or quality, as more reliable rankers disproportionately 
inform $\delta$.%
\footnote{This contrasts with the aggregation methods used by
\citet{hillebrecht2020community} and \citet{premand2020efficiency}, which assume that 
all rankers are equally reliable.
It is nevertheless possible that different individuals or groups generate rankings of
differing qualities.
For example, in \citet{premand2020efficiency}, the ranking exercises were conducted 
by committees of leaders and non-leaders.
Non-leaders may have better information about the living standards of community 
members than leaders, perhaps due to more frequent interactions or because community
members conceal relevant information from leaders.}

One can incorporate subjective beliefs about ranker qualities through the use of 
priors.
Again following \citet{li2016bayesian}, we specify a simple prior where $\omega_r$
can take on three values: 0.5, 1, and 2 corresponding to low-quality, mediocre,
and reliable rankers, respectively.
More specifically, we let
\begin{align}
    \omega_r &= \begin{cases} 0.5 &\text{with probability } \lambda_1^r \\
    1 &\text{with probability }  \lambda_2^r \\
    2 & \text{with probability } \lambda_3^r
    \end{cases}
\end{align}
where prior beliefs are incorporated through the choice of $\lambda_1$, $\lambda_2$, 
and $\lambda_3$ for each ranker. 
The full posterior for the multiple-ranker model is presented in Appendix 
\ref{appendixa}.
Once again, we use the samples from the posterior to generate the scores for
potential beneficiaries as $x_{ij} \hat{\delta}$.%
\footnote{In generating out-of-sample predictions, we set the random effects
equal to their expected value, which is zero in this case. 
As such, the random effects only influence the scores through their 
effect on $\delta$.}

\subsection{Elite capture}
\label{ssec: elite_capture}

As mentioned, CBT can be subject to elite capture whereby local elites influence the
beneficiary selection process in order to privilege relatives or friends.
Evidence of elite capture has been found in China \citep{han2019community},
Ethiopia \citep{caeyers2012political}, India \citep{besley2012just, 
panda2015political}, Malawi \citep{kilic2015decentralised, 
basurto2020decentralization}, and Tanzania \citep{pan2012vouching}, to name a few.
While there are instances where elite capture has been found to be negligible or 
non-existent \citep{alatas2012targeting, bardhan2010impact, schuring2014preferences},
where it is a potential threat to the targeting process, it may be useful to have 
a way to mitigate its influence.
We thus outline a method for removing the influence of elite connections on 
the scores for potential beneficiaries.

Let $x_{ik} = [x_{ik}^s, x_{ik}^e]$ where $x_{ik}^s$ is a row vector of 
sociodemographic characteristics and $x_{ik}^e$ is a row vector capturing 
elite connections for household $i$ in community $k$.
While there could be a variety of ways to specify $x_{ik}^e$, a simple specification 
could consist of a single indicator variable that captures whether or not any 
household member holds a leadership position or is related to anybody in a 
leadership position \citep{alatas2012targeting}.
We can then define the corresponding vector of coefficients as $\delta = [\delta^s,
\delta^e]$ and apply the models outlined in Eqs.\ (\ref{eq:simple}) or 
(\ref{eq:mrankers}) to retrieve estimates of $\delta$.
For well-specified models, the inclusion of $x_{ik}^e$ removes any confounding
influence of elite connections on $\delta^s$ and, as such, we can generate 
unbiased scores for potential beneficiaries as $x_{ij}^s \hat{\delta}^s$ 
(i.e., we set $\delta^e = 0$ when computing the scores).

\subsection{Auxiliary information}
\label{ssec: aux_info}

In the event that household survey data is available, one can use this auxiliary
data to improve the predictive accuracy of the model.
To this end, let $M \subseteq G$ represent the set of communities sampled for some
household survey and index these communities by $m$.
We assume that the household survey contains information on household incomes or
expenditures and denote the associated variable by $y_{im}$.
We further assume that the household survey includes sociodemographic
information for all households and denote the row vector of such characteristics
as $x_{im}$.
We then specify the following model:
\begin{equation}
y_{im} = \phi+ x_{im}\gamma + \psi_{im}    
\end{equation}
where $\phi$ is the intercept, $\gamma$ is a column vector of parameters, and 
$\psi_{im} \sim \text{N}(0, \sigma_\psi^2)$.
Note that the above model is equivalent to the standard form of PMT.

We can then specify a new (joint) likelihood as follows:
\begin{align}
P(z, y \mid \tilde{z}, \theta) &= P(z \mid \tilde{z}) P(y \mid \phi, \gamma,
\sigma_\psi^2) \\ \nonumber
&= \Bigg \{ \prod_k \prod_r \text{I}\big[\text{rank}(\tilde{z}_k^r) 
= z_{k}^r \big] \Bigg \} \Bigg \{\prod_m \prod_i \text{N}\left( y_{im} \mid 
\phi + x_{im}\gamma ,\sigma_\psi^2 \right) \Bigg \} 
\end{align}
where $\theta = \{\alpha, \delta, \omega, \phi, \gamma, \sigma_\psi \}$ denotes a 
vector of all fixed parameters.
The first component of the above is the likelihood of the ranked data and 
the second component is the likelihood associated with the auxiliary data.
The full posterior, including a description of all priors, is presented in
Appendix \ref{appendixa}.
A critical feature of this model is the priors we place on $\delta$
and $\gamma$.
In particular, we let $\delta, \gamma \sim \text{N}\left(\mu, \Sigma  \right)$
such that both parameter vectors are drawn from the same multivariate normal
distribution with mean vector $\mu$ and covariance matrix $\Sigma$.
The auxiliary data is thus permitted to inform $\mu$ and $\Sigma$, which in 
turn informs $\delta$.
Below we show that this procedure improves the predictive accuracy of the 
targeting scores $x_{ij}\hat{\delta}$ (or $x_{ij}^s\hat{\delta}^s$ when adjusting 
for elite capture).

\subsection{Dynamic updating}
\label{ssec: dynamic_updating}

Using data from Burkina Faso, \citet{hillebrecht2020dynamic} examined the dynamic
targeting performance of PMT and CBT in terms of identifying the consumption poor. 
While the authors found that PMT initially outperformed CBT, the performance of PMT
deteriorated faster than that of CBT, to the extent that CBT outperformed PMT 
after just one year.
One reason for the dynamic success of CBT was that communities implicitly weighted
sociodemographic characteristics in a way that better predicted future poverty.
Though our method might be expected to inherit the forward-looking nature of 
community rankings, the fact that both PMT and CBT were subject to non-negligible
deterioration suggests that targeting mechanisms need to be regularly updated.

While many poverty-alleviation programs using PMT conduct periodic updates
\citep{kidd2011, kidd2017}, the procedure for updating the PMT algorithm generally
neglects important information.
In particular, re-estimating PMT weights by applying frequentist estimation methods
(e.g., OLS) to new data implicitly assumes uniform priors on all weights, when in 
fact relevant information is available from previous implementations.
The implication of this ``memoryless'' updating is that it sacrifices predictive
accuracy.
In contrast, our Bayesian framework naturally accommodates updating, as the 
posterior from any given period can be used as the prior in some subsequent
period.
The Bayesian approach thus provides a coherent way to dynamically update the
targeting algorithm as new information becomes available.

To illustrate, let $P(\theta \mid z_{1:t+1})$ denote the posterior distribution
of the model's parameters $\theta$ conditional on all data available up to some time 
period $t+1$.
We can then show that the posterior $P(\theta \mid z_{1:t+1})$ is proportional to the
likelihood of some new data $P(z_{t+1} \mid \theta)$ multiplied by the posterior from 
the previous period $P(\theta \mid z_{1:t})$:
\begin{align}
    P(\theta \mid z_{1:t+1})&  \propto P(z_{1:t+1} \mid \theta)P(\theta)  \\ \nonumber
    &= P(z_{t+1} \mid \theta) P(z_{1:t} \mid \theta)P(\theta)  \\ \nonumber
    &\propto P(z_{t+1} \mid \theta) P(\theta \mid z_{1:t}).
\end{align}
That is, using the posterior from a previous period as the prior in a subsequent
period is mathematically equivalent to conducting the analysis using data from
all periods.
The above assumes that all data contribute equally relevant information, which may
be questionable given the results of \citet{hillebrecht2020dynamic}.
To overweight more recent information, one can simply diffuse $P(\theta \mid z_{1:t})$
around the first moments, thus weakening the influence of previous data on the 
posterior.
See Appendix \ref{appendixa} for additional discussion.


\section{Data}
\label{sec: data}

We illustrate our hybrid targeting framework using data from Burkina Faso and 
Indonesia.
Our primary goal is to understand the out-of-sample predictive performance of
our model and, as such, we split each dataset into three mutually exclusive
samples: (1) a test sample that we use to evaluate the performance of the model,
(2) a training sample that we use to estimate the model, and (3) an auxiliary sample 
that we reserve for augmenting the training sample to estimate the model that 
incorporates auxiliary information (see Section \ref{ssec: aux_info}).
In what follows, we discuss the data from both countries and outline our 
approach for splitting each sample.

\subsection{Burkina Faso}
\label{ssec: burkina}

\citet{hillebrecht2020community} examined the performance of alternative 
targeting methods using data from the department of Nouna in the northwest
of Burkina Faso.%
\footnote{The data is not publicly available and was provided to us by the 
authors.}
With the objective of increasing enrollment rates in a community-based health 
insurance scheme, the Burkinab\'{e} Ministry of Health in partnership with a
local NGO offered a 50 percent discount on the premium for the poorest
20 percent of households in a number of villages and urban neighborhoods
in Nouna.
To identify beneficiaries, each village or neighborhood conducted CBT exercises 
in the years 2007, 2009, and 2011.
\citet{hillebrecht2020community} focused on the 2009 campaign where CBT
exercises took place in 58 different communities, including 36 villages
and 22 neighborhoods in Nouna Town.

The CBT exercises began with a public meeting that included focus group 
discussions and an election of three local informants that were charged with 
ranking all community members in terms of the poverty and wealth criteria
identified by the focus groups.
The three local informants were then physically separated from the assembly to
complete the ranking exercises.
The final beneficiaries were then selected based on agreements across
the three rank orderings.
First, all households ranked among the poorest 20 percent by all informants 
were automatically eligible.
Second, all households ranked among the poorest 20 percent by two informants were
deemed eligible, unless the number of households exceeded the quota, in which 
case the informants were consulted to narrow the list.
Finally, if necessary, the informants were consulted to select the remaining
households from those that were ranked among the poorest 20 percent by one
informant.

The Nouna Household Survey was also conducted in 2009 and it includes 
information on a sample of 655 households in the 58 communities where the
CBT exercises were conducted. 
Due to some missing information on key variables, we only use 608 observations
from this dataset.
Merging the household survey with the rankings from the CBT exercises, we
then have information on the rank, monthly per capita consumption, and 
sociodemographic characteristics (e.g., household demographics,
dwelling characteristics, and assets) of each household.
We split this sample equally into our testing and training samples.
For our auxiliary sample, we use data from the 2007 round of the Nouna
Household Survey (654 households), which is the same sample that 
\citet{hillebrecht2020community} used to calibrate their PMT model.
The reader is referred to their paper for more detailed discussion.

\subsection{Indonesia}
\label{ssec: indonesia}

\citet{alatas2012targeting} conducted a targeting experiment in three provinces
of Indonesia -- North Sumatra, South Sulawesi, and Central Java -- that were 
selected to be geographically and ethnically representative of the country.%
\footnote{The data is publicly available on the Harvard Dataverse 
\citep{alatas2013data}.}
The researchers randomly selected 640 villages across the three provinces and 
then selected one subvillage in each to participate in the experiment.
The subvillages contained 54 households on average.
Subvillages were then randomly assigned to three ``treatment'' arms, each of which 
entailed a different procedure for selecting beneficiaries to receive a one-time 
cash payment of Rp.\ 30,000 (approximately \$3 and roughly equal to the daily wage 
for the average laborer).
While the beneficiary quota was different for each subvillage, on average 
about 30 percent of households were selected.

The treatments assigned to the subvillages included PMT, CBT, and a hybrid 
approach where CBT was used to select 1.5 times the quota of beneficiaries
and then the list was narrowed using PMT.
Both the CBT and hybrid subvillages thus conducted full community ranking 
exercises where residents met to rank households from ``poorest'' to ``most
well-off.''
In these exercises, a facilitator presented community members with a randomly
ordered set of index cards, each of which displayed the name of one household.
Through public discussions, the community members then ordered the households
by placing the index cards on a string hung from a wall.
The lowest-ranking households were then selected to receive benefits according
to the predetermined quota for each subvillage.

Prior to the targeting exercises in late 2008, an independent survey firm
gathered data on a number of households from each subvillage, including
the subvillage head and a random sample of eight households.
The sample consists of a total of 5,756 households, of which we use 5,004
due to missing information for some variables.
The targeting exercises were conducted shortly after the data were collected
and the resulting household ranks were recorded for each household.
In addition to the rank for each household, we observe monthly expenditure per 
capita (Rp.\ 1000s) and rich sociodemographic information, including 
household demographics, dwelling characteristics, and assets.
For additional details on the dataset, the reader is referred to
\citet{alatas2012targeting}.

Regarding splitting the dataset, we take all households that conducted 
community ranking exercises -- namely, the 3,375 households from the CBT and 
hybrid communities -- and split them equally into training and testing samples.
Note that our interest is in predicting community rankings, so for the hybrid
communities we discard the rankings generated by the hybrid 
procedure and focus on the underlying rankings from the community targeting 
exercises.
Finally, for the auxiliary sample, we use the households assigned to the PMT
treatment (1,629 households).
The subvillages assigned to the PMT treatment did not conduct community ranking
exercises, so we do not have the necessary information on these households to 
include them into either the testing or training samples.


\section{Results}
\label{sec: results}

In this section, we present the results from our analysis of the data from Burkina 
Faso and Indonesia.%
\footnote{All data and code used for our empirical illustration can be found at
\url{https://github.com/LendieFollett/Hybrid-Targeting}.}
We first present some baseline results that focus on comparing our hybrid model to 
competing statistical targeting methods, namely PMT and the benchmark probit model.
We then consider in more detail our various model extensions, including our
generalization to multiple rankers, our adjustments for elite capture, the 
inclusion of auxiliary information, and dynamic updating.

\subsection{Baseline results}
\label{ssec: results_baseline}

\begin{figure}[!t]
\begin{center}
\includegraphics[scale=0.4]{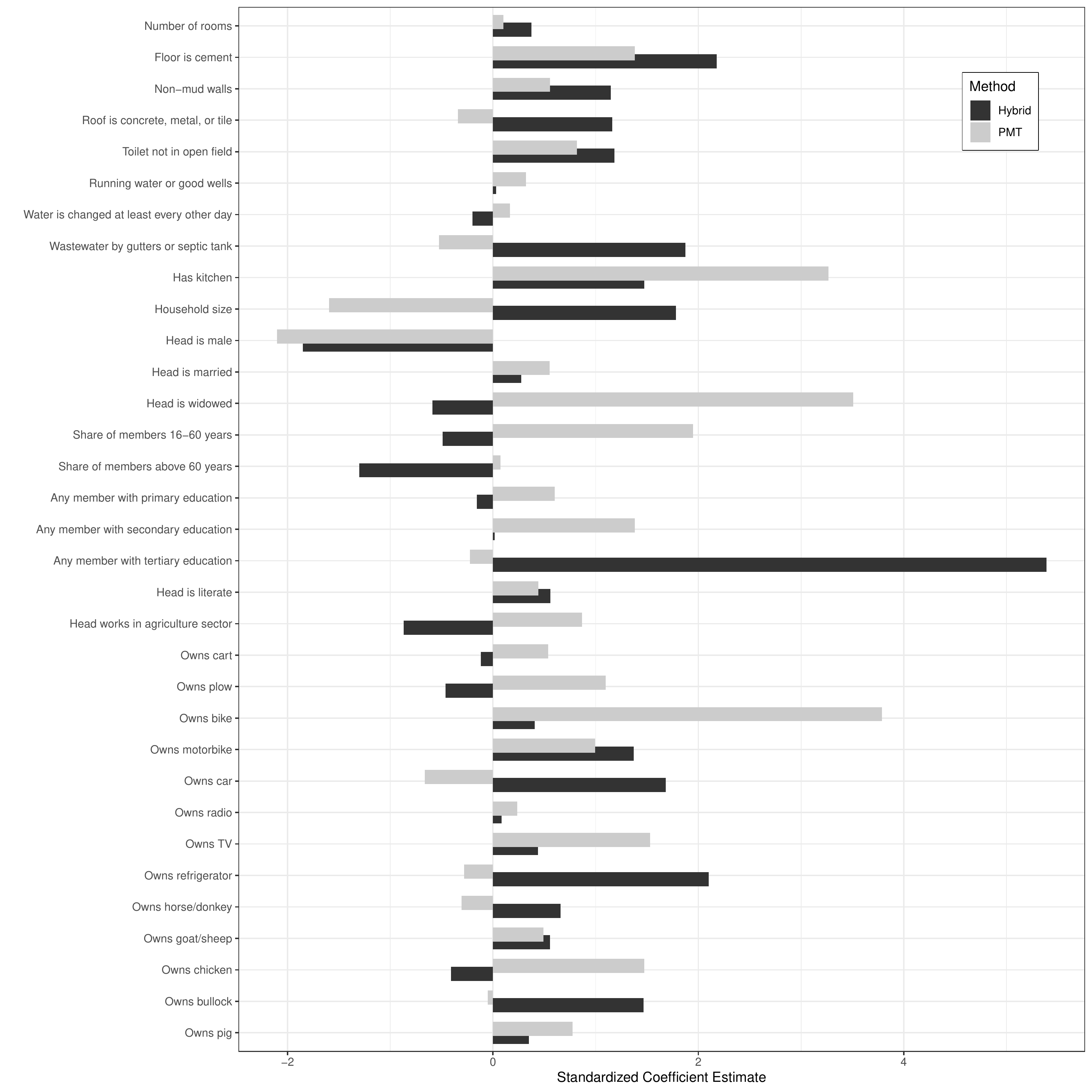}
\caption{Standardized coefficient estimates (Burkina Faso)}
\label{fig:coef_hillebrecht}
\end{center}
\end{figure}

\begin{figure}[!t]
\begin{center}
\includegraphics[scale=0.4]{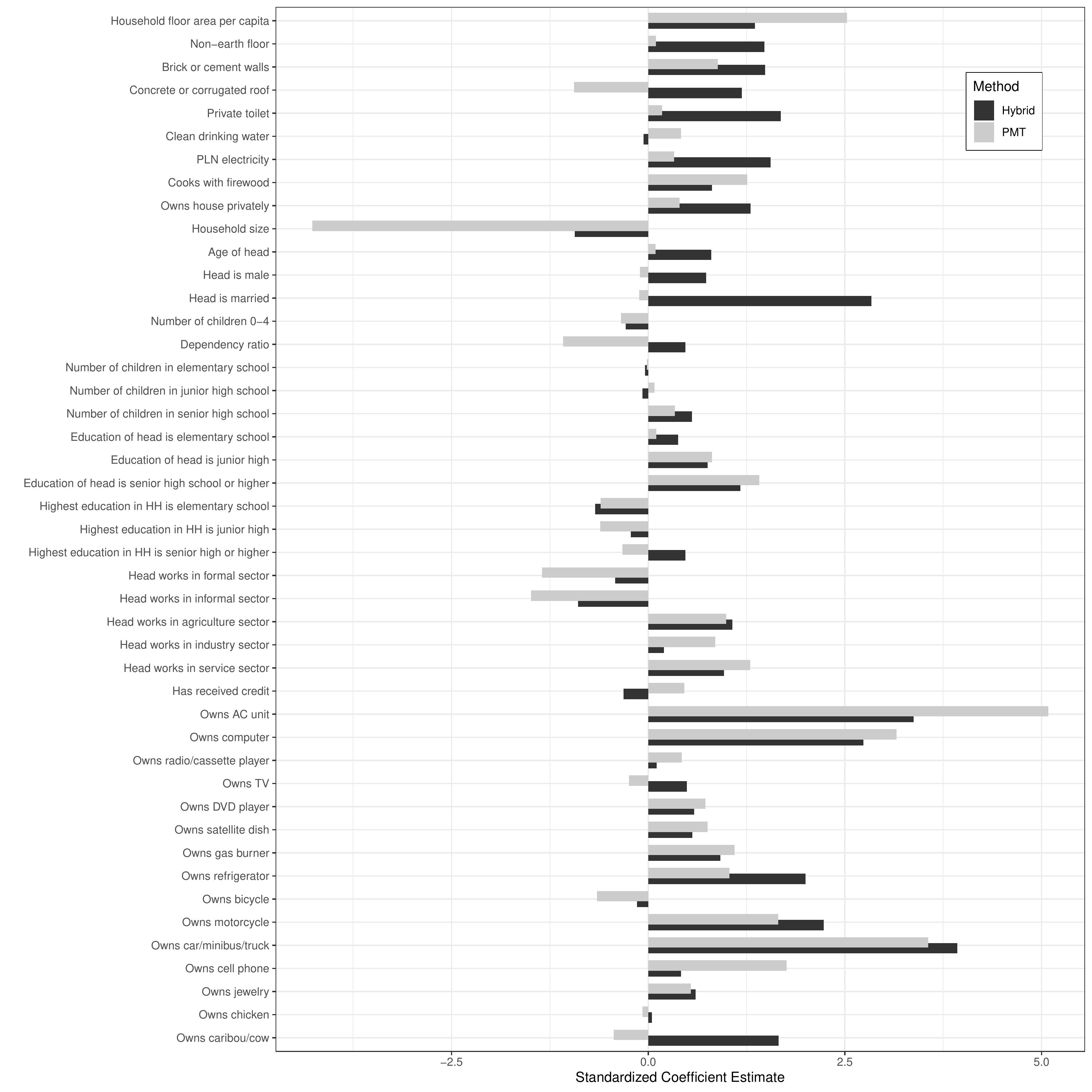}
\caption{Standardized coefficient estimates (Indonesia)}
\label{fig:coef_alatas}
\end{center}
\end{figure}

We have argued that PMT neglects local preferences by imposing an income- or
expenditure-oriented definition of poverty on communities.
To illustrate the implications of this issue, Figures \ref{fig:coef_hillebrecht} and
\ref{fig:coef_alatas} provide a representation of how our hybrid model weights
sociodemographic characteristics relative to PMT.
For the Burkinab\'{e} data (Figure \ref{fig:coef_hillebrecht}), we apply the hybrid model
outlined in Section \ref{ssec: multiple_rankings} because this dataset includes 
information from multiple rankers.
For the Indonesian data (Figure \ref{fig:coef_alatas}), we use our most basic hybrid
model outlined in Section \ref{ssec: basic_model}.
We use all available observations to estimate each model and the included covariates
are taken directly from the specifications used in \citet{hillebrecht2020community}
and \citet{alatas2012targeting} for the Burkinab\'{e} and Indonesian data, 
respectively.%
\footnote{More specifically, we use all available observations to estimate our hybrid
model and then estimate the PMT model using the same sample.
The sociodemographic characteristics used by \citet{hillebrecht2020community} 
are listed in the repository provided to us by the authors.
We have added the variable household size to their specification.
The characteristics used by \citet{alatas2012targeting} are presented
in their Table 11.
For ease of interpretation, we omit a small number of quadratic and interaction
terms from their specification.}

To compare coefficients across the hybrid and PMT models, we standardize 
the estimated coefficients by dividing each by the average of the absolute value of 
the coefficients from each model. 
With this standardization, the sign of each effect reflects the sign of the 
associated coefficient whereas the magnitude captures the average marginal rate
of substitution (MRS).
The average MRS tells us how many units, on average, each covariate must change to 
compensate for a one-unit increase in the covariate of interest while holding the 
outcome constant.
We view the average MRS as a measure of variable importance, reflecting the 
relative sensitivity of the model to a given covariate.
This interpretation requires that all covariates are placed on a similar scale and we
accomplish this by dividing all continuous variables by two times their standard 
deviation \citep{gelman2008scaling}.
See Appendix \ref{appendixb} for detailed discussion.

Figures \ref{fig:coef_hillebrecht} and \ref{fig:coef_alatas} show that communities 
implicitly weight sociodemographic characteristics quite differently from PMT.
Regarding dwelling characteristics, we see some stark contrasts, particularly
with respect to roof type.
For both Burkina Faso (``roof is concrete, metal, or tile'') and Indonesia 
(``concrete or corrugated roof''), we find that our model positively weights
roofing type whereas PMT counterintuitvely estimates negative weights.
As dwelling characteristics are easily observable indicators of living standards,
it is natural that communities give higher ranks or scores to households with
higher-quality dwellings.%
\footnote{Recall that those households with the lowest ranks or scores are selected
for program inclusion.}
In the case of roof type, PMT thus appears to be at odds with such preferences
by prioritizing households with high-quality roofing, all else equal. 
We view this results as a fairly clear illustration of how PMT can be 
inconsistent with local preferences.

In terms of demographic characteristics, we find some pronounced differences in
how communities weight education.
For Burkina Faso, the hybrid model indicates that the average MRS for ``any member 
with tertiary education'' is 5.14, meaning that a one-unit increase in this 
variable must be compensated for by a relatively large 5.14 unit change in the 
other covariates on average.
In contrast, PMT is associated with a relatively small negative effect of -0.22.
We similarly find some notable disagreements related to advanced education in the
Indonesian data where the effects on ``highest education in HH is senior high
or higher'' also differ in sign across the two models.
Regarding the other demographic variables, it is worth highlighting the differing
weights placed on marital status, namely ``head is widowed'' in the Burkinab\'{e}
results and ``head is married'' in the Indonesian results.
The PMT results here once again appear inconsistent with local preferences.

Figures \ref{fig:coef_hillebrecht} and \ref{fig:coef_alatas} also
show some notable differences in how communities weight asset variables.
Most interestingly, we find that communities implicitly disagree with the PMT 
weights on some livestock-related variables, namely ``owns bullock'' in the 
Burkinab\'{e} data and ``owns caribou/cow'' in the Indonesian data.
For example, in the Burkinab\'{e} data, the effect on ``owns bullock'' from
the hybrid model is 1.49 whereas that from the PMT model is -0.05.
While the effect magnitudes are quite different, it is worth emphasizing
that PMT places a negative weight on livestock ownership in this case, which
is surprising given that livestock are important for income-generating activities.
The Indonesian data show a similar result.
Taken together, the results from Figures \ref{fig:coef_hillebrecht} and 
\ref{fig:coef_alatas} thus show that PMT conflicts with community targeting 
preferences, particularly with respect to important traits like roofing type, 
education, and livestock ownership.

\begin{figure}[!t]
\begin{subfigure}{.5\textwidth}
  \centering
  \includegraphics[scale=0.4]{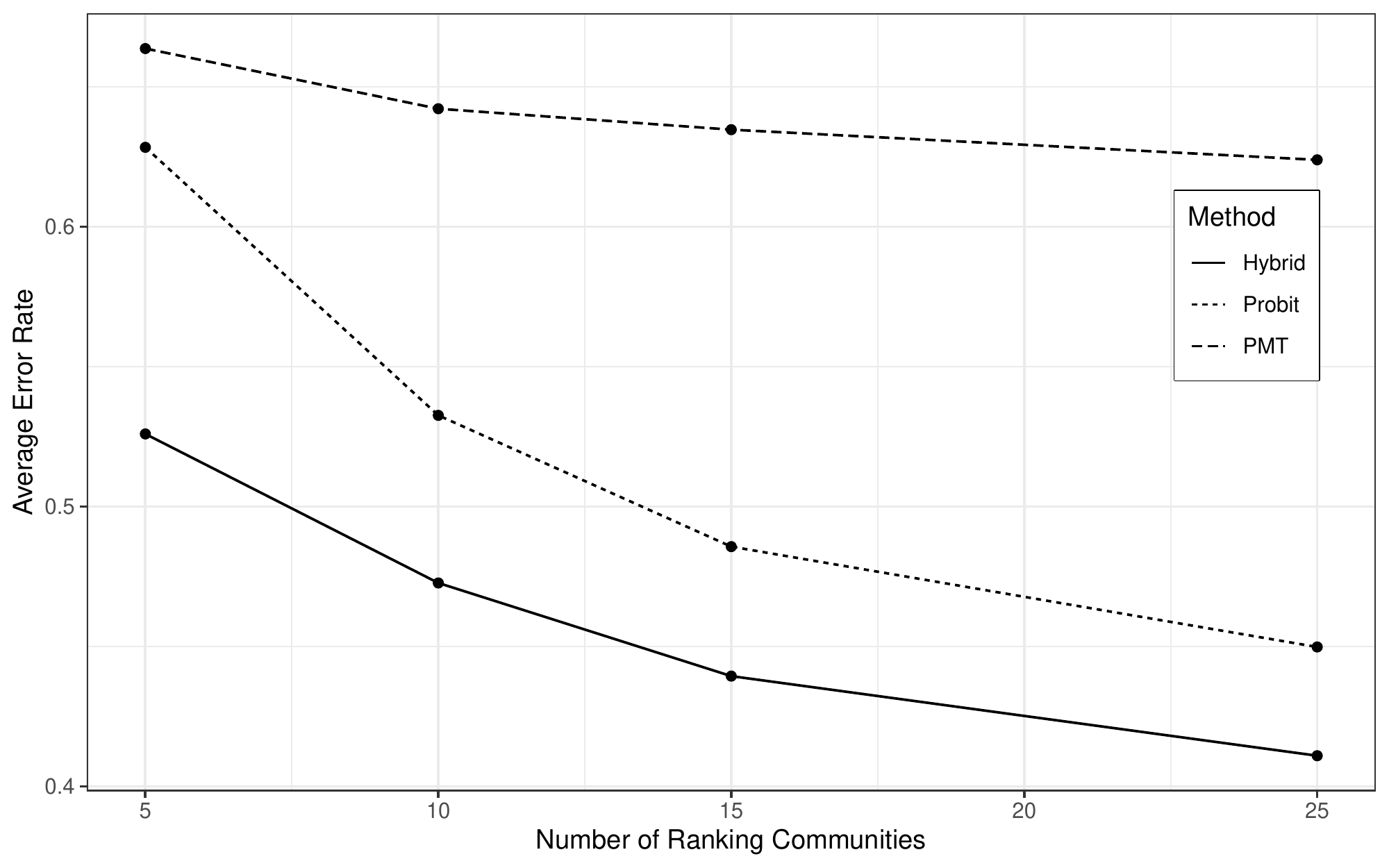}  
  \caption{Burkina Faso}
\end{subfigure}
\begin{subfigure}{.5\textwidth}
  \centering
  \includegraphics[scale=0.4]{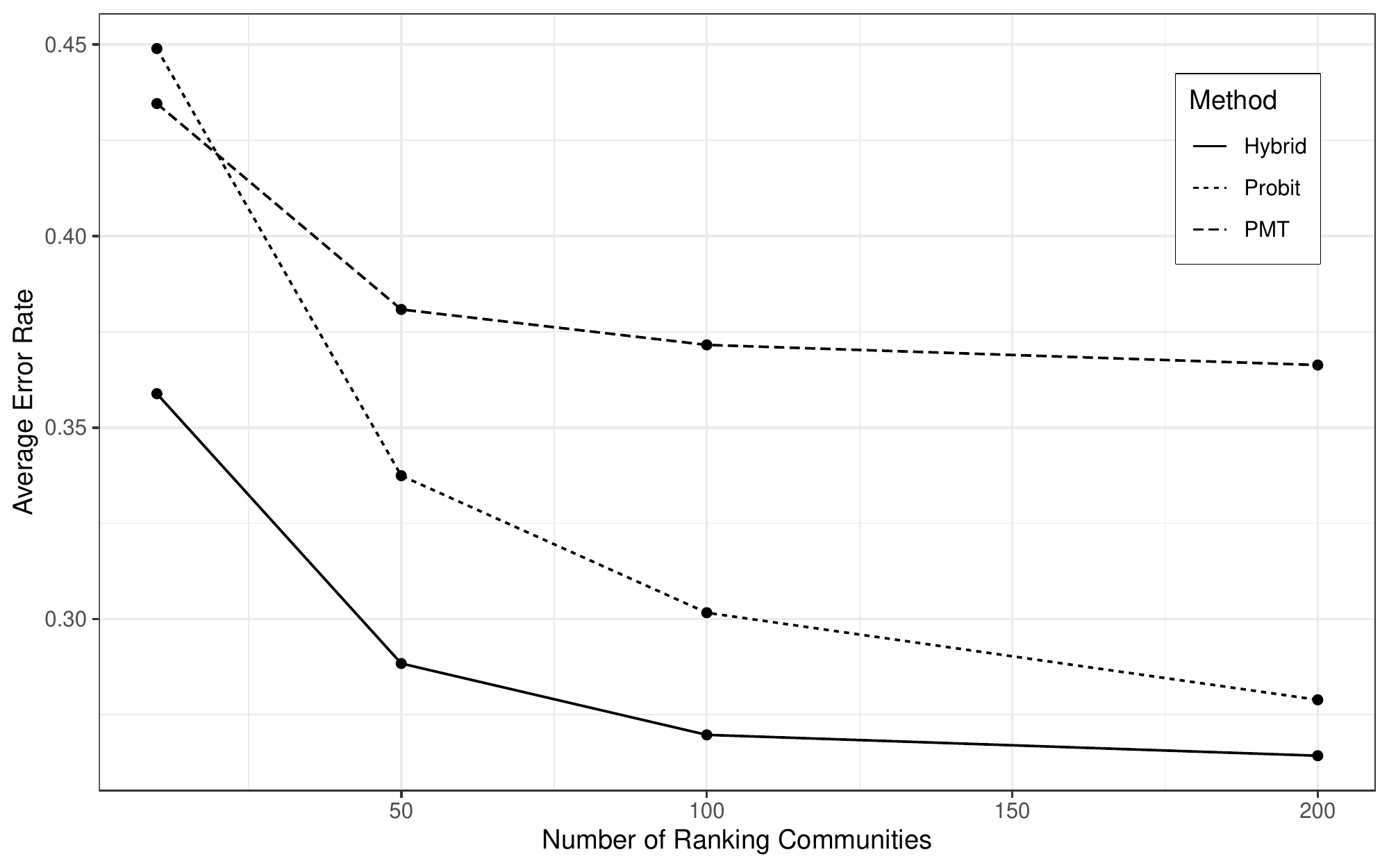}  
  \caption{Indonesia}
\end{subfigure}
\caption{Out-of-sample predictive performance of select models}
\label{fig:er_hybrid}
\end{figure}

Figure \ref{fig:er_hybrid} presents results related to the out-of-sample 
predictive performance of our model where the estimand of interest is
community preference rankings.
We measure performance through exclusion and inclusion error rates
\citep{coady2004targeting, stoeffler2016reaching, brown2018}. 
The exclusion error rate is defined as the share of the truly poor
(i.e., the households actually selected by the communities) not selected for 
program participation by any given statistical targeting method.
The inclusion error rate is defined as the share of households selected by a 
given targeting method that are not truly poor (i.e., those households not selected
by the communities).
We set the number of households selected by any method to be equal to the number of 
truly poor and this implies that the exclusion and inclusion error rates will be 
identical.%
\footnote{More specifically, we set the number of households selected from any 
community to be equal to the number of truly poor households in that community.
Following from the discussion in Section \ref{sec: data}, we then select 
approximately 20 percent of households when using the data from Burkina Faso and 
30 percent of households when using the Indonesian data.}
We will thus simply refer to the error rate in what follows.

Panel (a) in Figure \ref{fig:er_hybrid} uses the data from Burkina Faso and 
displays average error rates across 30 replications for varying numbers
of communities sampled from the training data (see Section \ref{sec: data} for 
discussion of our training-test splits).
That is, for a given number of training communities, we repeatedly sample 
communities from the training sample and for each replication we use the
estimated model to calculate error rates in the test sample.
We conduct this procedure for three models: our hybrid model for multiple rankers, 
the benchmark probit, and PMT.
Panel (b) conducts the same exercise, but uses the data from Indonesia.
Note that the Indonesian data uses a single ranking scheme, so here we use
the basic hybrid model outlined in Section \ref{ssec: basic_model}.
Further note that we use different community numbers across the two applications
because of the differing sample sizes.%
\footnote{Recall from Section \ref{sec: data} that the Burkinab\'{e} data contains
roughly 11 households per community and the Indonesian data contains about nine 
households per community.
A simple way to translate communities sampled into households sampled is then to 
multiply the number of communities sampled by 10.}

The expected error rate under purely random targeting is 80 percent for Burkina Faso 
and 70 percent for Indonesia.%
\footnote{Once again, the Burkinab\'{e} community ranking exercises selected
the poorest 20 percent of households and the Indonesian exercises selected
the poorest 30 percent.
A purely random targeting method that selects the same proportion of households
will achieve an expected error rate of one minus that proportion.}
All the methods presented in Figure \ref{fig:er_hybrid} thus outperform
purely random targeting in all cases, at least on average.
Given our previous results showing that communities implicitly weight 
sociodemographic traits differently from PMT, it is unsurprising that PMT 
performs relatively poorly across all sample sizes.
While the simple probit model outperforms PMT in the vast majority of cases, it 
never outperforms our hybrid model.
Recalling our discussion in Section \ref{ssec: probit}, this is also an expected 
result because the probit model is subject to a number of limitations (e.g., loss 
of information by dichotomizing the community rankings).
One conclusion from Figure \ref{fig:er_hybrid} is then that, on average, our hybrid 
model outperforms random targeting, PMT, and the probit model for all sample sizes.

Consider in more detail the error rates achieved by the hybrid model.
For Burkina Faso (Indonesia), we find error rates ranging from 0.53 to 0.41
(0.36 to 0.26) for the smallest and largest sample sizes, respectively. 
While we have seen that the hybrid model outperforms PMT in terms of predicting
community rankings, we can also consider how well the hybrid model performs relative
to PMT's ability to predict household expenditures.
To this end, we have conducted similar experiments with both datasets to calculate
the error rates achieved by PMT in terms of predicting expenditures out of sample.%
\footnote{That is, for varying training sample sizes, we fit an OLS model of (log) 
expenditure per capita on all sociodemographic characteristics and then use the 
estimates to calcuate out-of-sample error rates.
A household's true poverty status in this experiment is determined by their 
expenditure per capita rather than their rank from the community exercises.
Given that these experiments are less computationally demanding, we use 1,000
replications for each sample size.}
For Burkina Faso (Indonesia), we find that PMT achieves error rates ranging from 
0.65 to 0.57 (0.40 to 0.30) for the smallest and largest sample sizes, respectively. 
The hybrid model thus also outperforms PMT on its own terms, with error rate
differences of 12-16 percentage points for Burkina Faso and about four 
percentage points for Indonesia.

\subsection{Model extensions}
\label{ssec: extensions_results}

As discussed in Section \ref{ssec: multiple_rankings}, our model not only 
accommodates multiple rankers, but also aggregates alternative ranking schemes 
according to their relative precision or quality. 
While the data from Burkina Faso includes multiple rankers, there is no fundamental
heterogeneity across the rankers given that each community simply selected three
informants to complete the ranking exercises.
Other implementations of community ranking exercises, however, have featured
distinct ranker types, which we might expect to differ systematically in terms of 
quality.
Recall the exercises conducted in Niger that consisted of three different 
ranker types: local leaders, non-leader women, and a mixed group of non-leaders
\citep{premand2020efficiency}.
To show how our multiple-ranker model adjusts for ranker quality, we conduct a 
simple experiment with the Burkinab\'{e} data where we artificially create a
non-informative ranker by randomly shuffling one ranking scheme from each community.

\begin{figure}[!t]
\begin{subfigure}{.5\textwidth}
  \centering
  \includegraphics[scale=0.3]{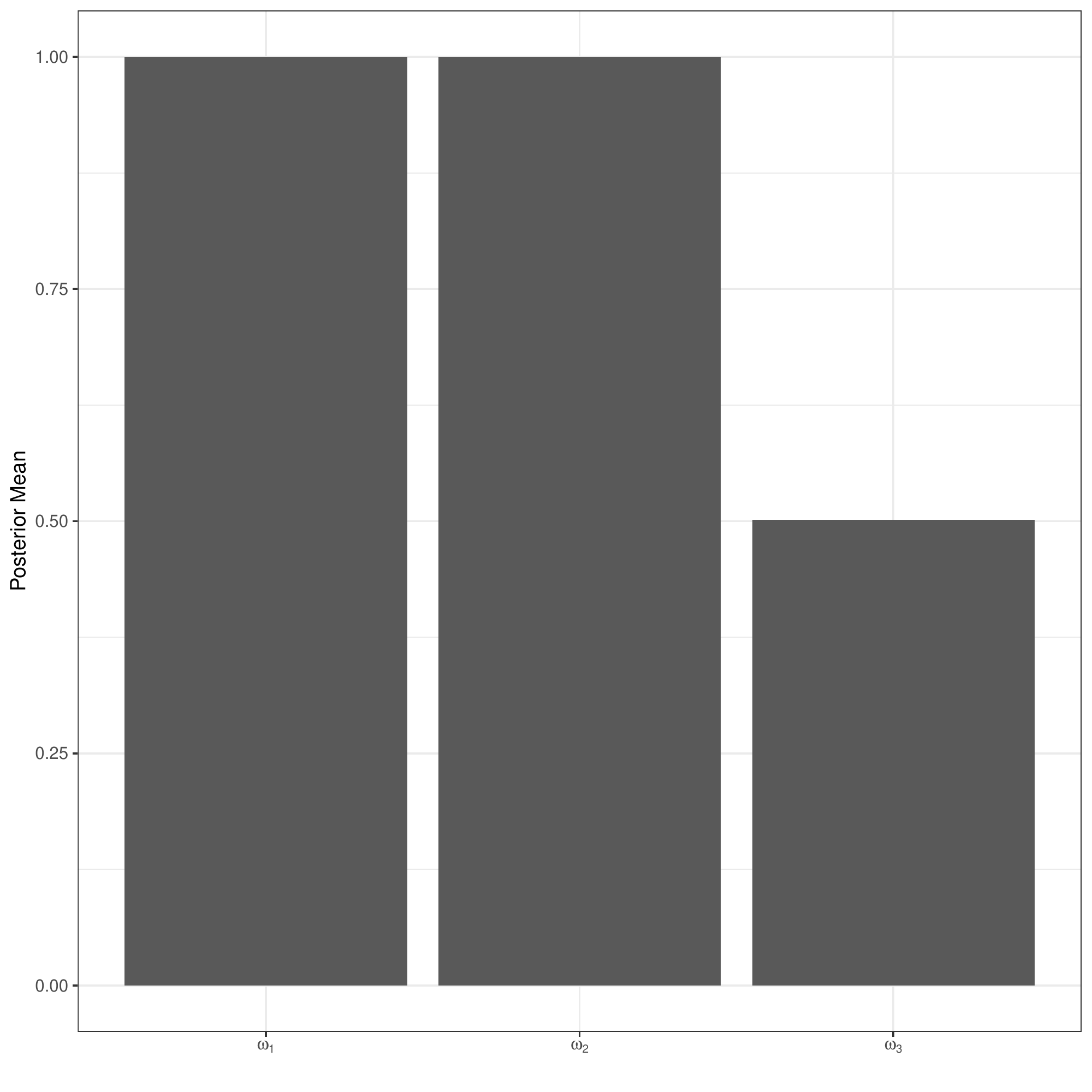}  
  \caption{Precision}
\end{subfigure}
\begin{subfigure}{.5\textwidth}
  \centering
  \includegraphics[scale=0.3]{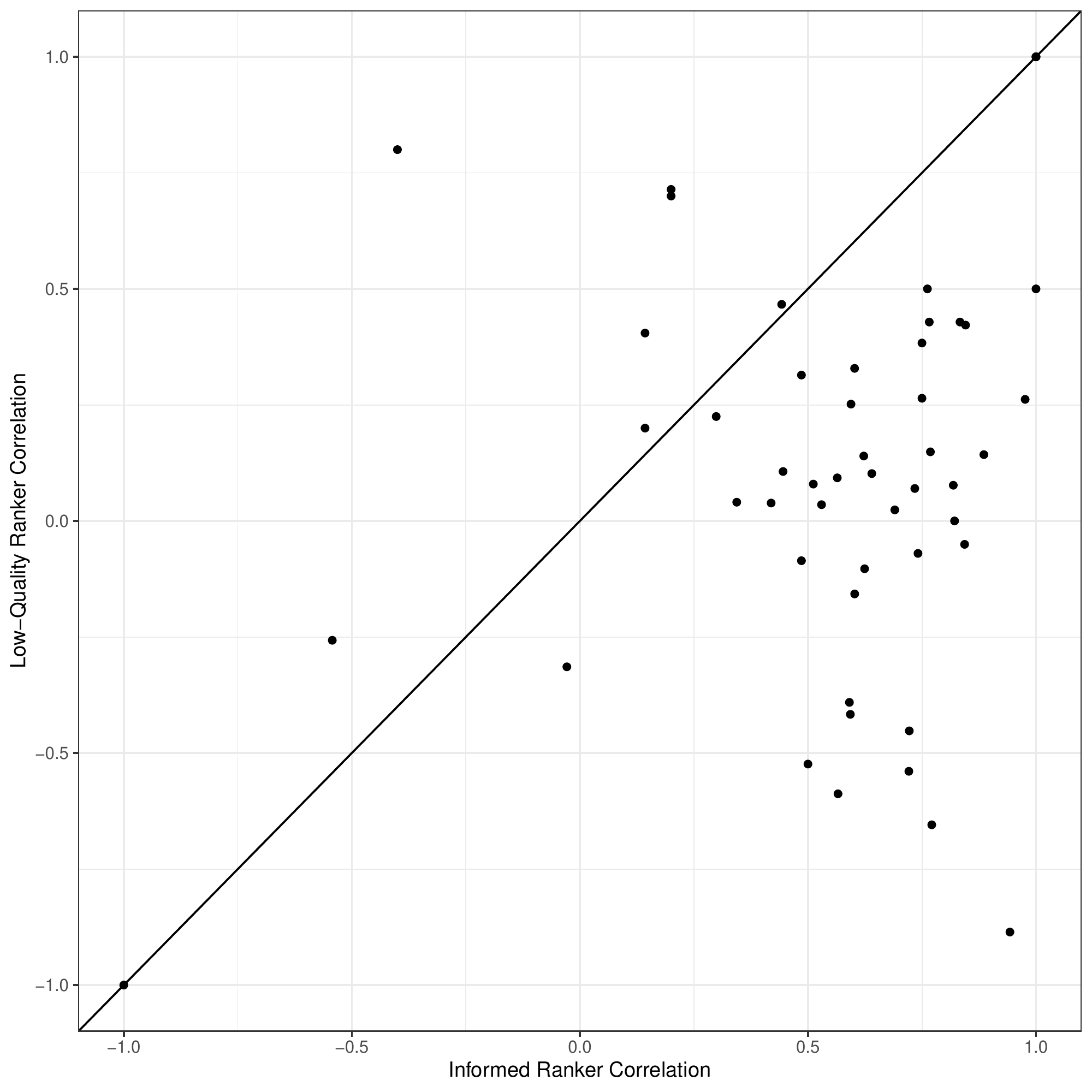}  
  \caption{Correlations}
\end{subfigure}
\caption{Heterogeneous rankers and the hybrid model}
\label{fig: multiple_rankers}
\end{figure}

The results from this exercise are presented in Figure \ref{fig: multiple_rankers}
and correspond to one run of the model using all available data.
Panel (a) plots the posterior means of each ranker's precision $\omega$ where
higher values indicate higher-quality ranking schemes.
Recalling that $\omega$ can take on three values -- 0.5, 1, and 2 for low-quality,
mediocre, and reliable rankers, respectively -- we see that the precision for the
shuffled ranker ($\omega_3$) clearly indicates low quality.
Panel (b) illustrates the implications of this adjustment: For each community
we calculate the rank correlation between the aggregated ranking generated by
our model and the rankings of the non-informative ranker and one informative
ranker.
Plotting the correlations associated with the non-informative ranker on that
for the informative ranker for each community, we see that the vast majority
of points fall below the diagonal.
Higher-quality rankers are thus more heavily weighted by our model.

Turning to our adjustments for elite capture, the Indonesian data includes a
variable for elite connections. 
This binary variable takes on the value of one for a household if (1) 
any member of the household held a formal leadership position in the village,
(2) at least two other households in the village identified the household as
having a member that held a formal or an informal leadership position, or
(3) the household was connected by blood or marriage to any household
meeting the first two conditions \citep{alatas2012targeting}.
In accordance with Section \ref{ssec: elite_capture}, we augment our basic 
hybrid model with this elite connections variable and then set the coefficient
to zero when calculating the targeting scores.
To reiterate, the purpose of this procedure is to remove any confounding
influence of elite connections from the targeting scores.

\begin{figure}[!t]
\begin{center}
\includegraphics[scale=0.4]{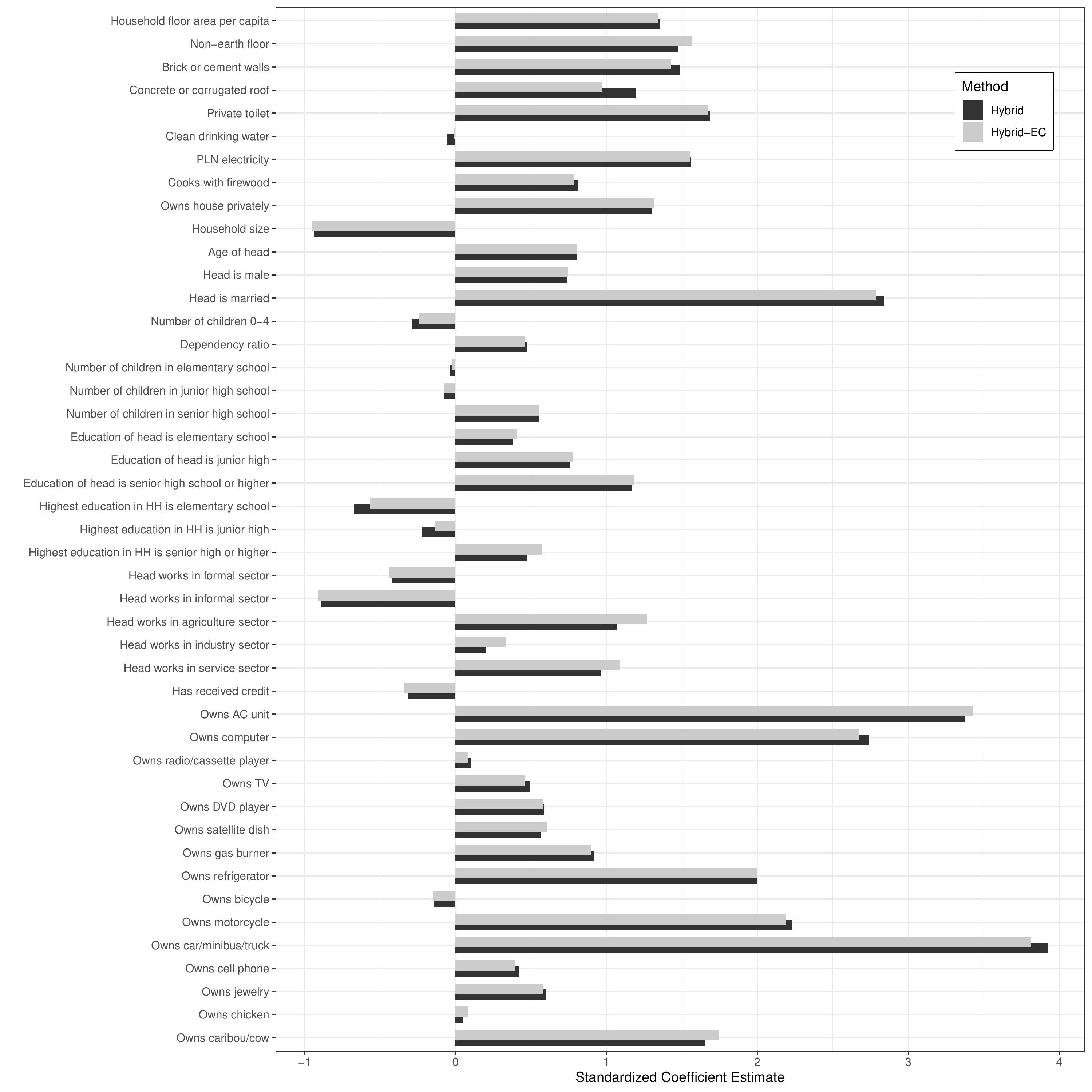}
\caption{Standardized coefficient estimates (elite capture)}
\label{fig: coef_EC}
\end{center}
\end{figure}

Estimating the model with all observations gives a standardized coefficient 
estimate on the elite connections variable of -0.04 with a 95 percent credible
interval of -0.46 to 0.34.
While the effect size is relatively small and imprecisely estimated, the negative 
sign suggests some tendency for connected households to be privileged in the 
community ranking exercises, all else equal.%
\footnote{This result contrasts with \citet{alatas2012targeting}, who find that
connected households are less likely to be treated (see their Table 7).
The primary reason for our differing results is that we use the rankings as our
outcome variable whereas Alatas et al.\ use a dichotomized ranking variable.
Specifically, we find a qualitatively similar result to Alatas et al.\ when
we regress a dichotomized ranking variable on elite connections (and all other
covariates) using our sample.
This suggests that the treatment of the outcome variable explains the differing
results.}
To see how the addition of the elite connections variable affects the coefficients
on the remaining covariates, Figure \ref{fig: coef_EC} plots the standardized 
coefficients from the hybrid model with (``Hybrid-EC'') and without (``Hybrid'') the 
elite capture correction.
Note the subtle differences regarding the occupational variables, namely
those related to whether the household head works in the industrial, service, or
agricultural sector.
We find that omitting the elite connections variable imparts a downward bias
on these coefficients, presumably because connected households are more likely
to be employed.%
\footnote{The relationship between elite connections and employment in any sector
follows from the fact that we find a negative relationship between elite connections 
and a household's ranking.
That is, the relationship between elite connections and employment in any sector
must be positive to impart a downward bias on the coefficients.}

\begin{figure}[!t]
\centering
\includegraphics[scale=0.4]{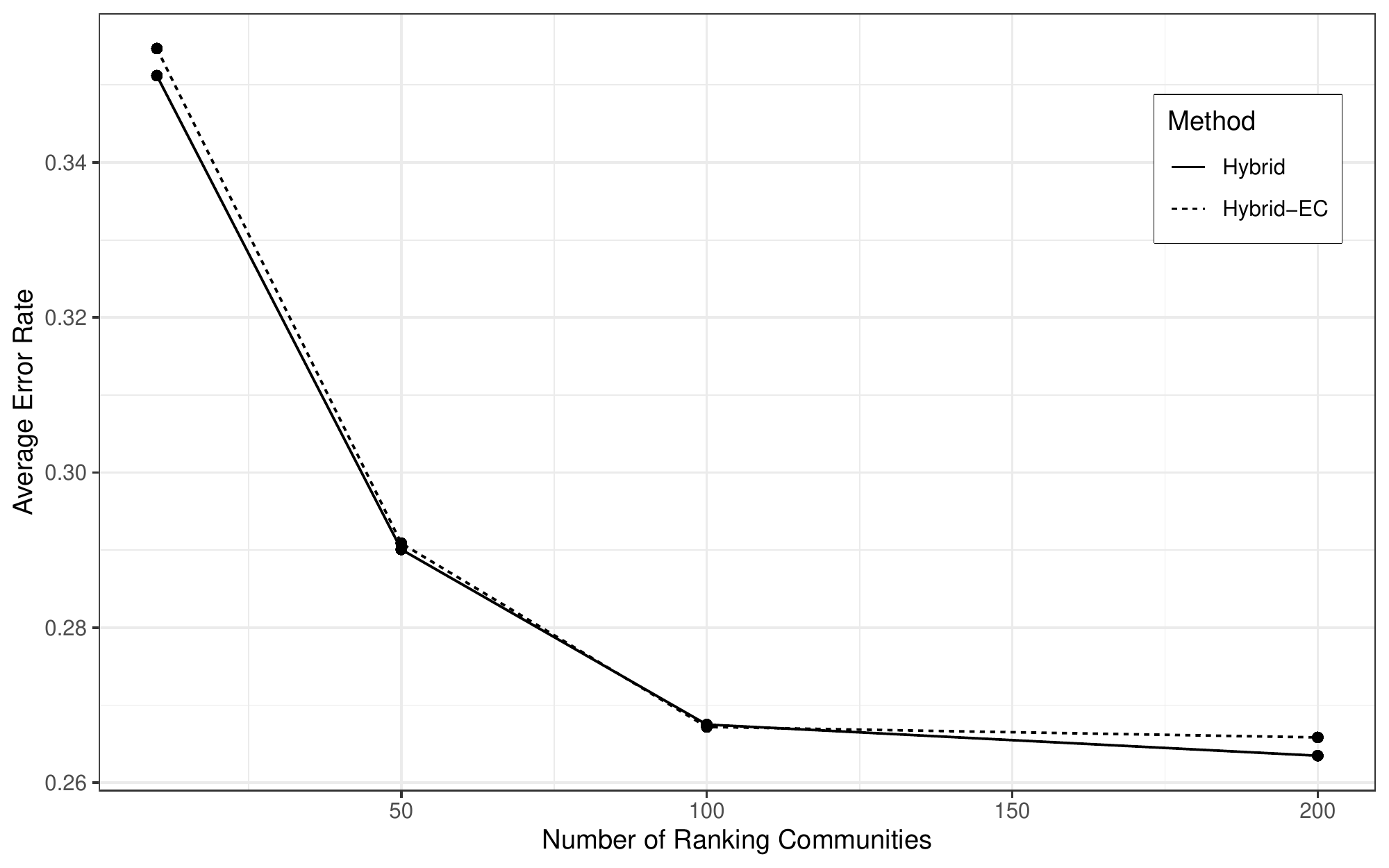}  
\caption{Out-of-sample predictive performance with elite capture correction}
\label{fig: er_EC}
\end{figure}

While there are evidently some coefficient changes after introducing the elite 
connections variable, there are few sign changes and the effect sizes remain
similar for most covariates.
This suggests that the out-of-sample predictive performance of the two models 
should be similar.
Figure \ref{fig: er_EC} displays average error rates across 30 replications,
once again varying the number of communities sampled from the training data.
In comparing the hybrid model with and without the elite capture correction,
we find the expected result that the average error rates are similar across
the two models.
Note, however, that the corrected model performs slightly worse than the 
uncorrected model for some sample sizes.
This is also an expected result given that the corrected model is constrained
by setting to zero the coefficient on the elite connections variable.

\begin{figure}[!t]
\begin{subfigure}{.5\textwidth}
  \centering
  \includegraphics[scale=0.4]{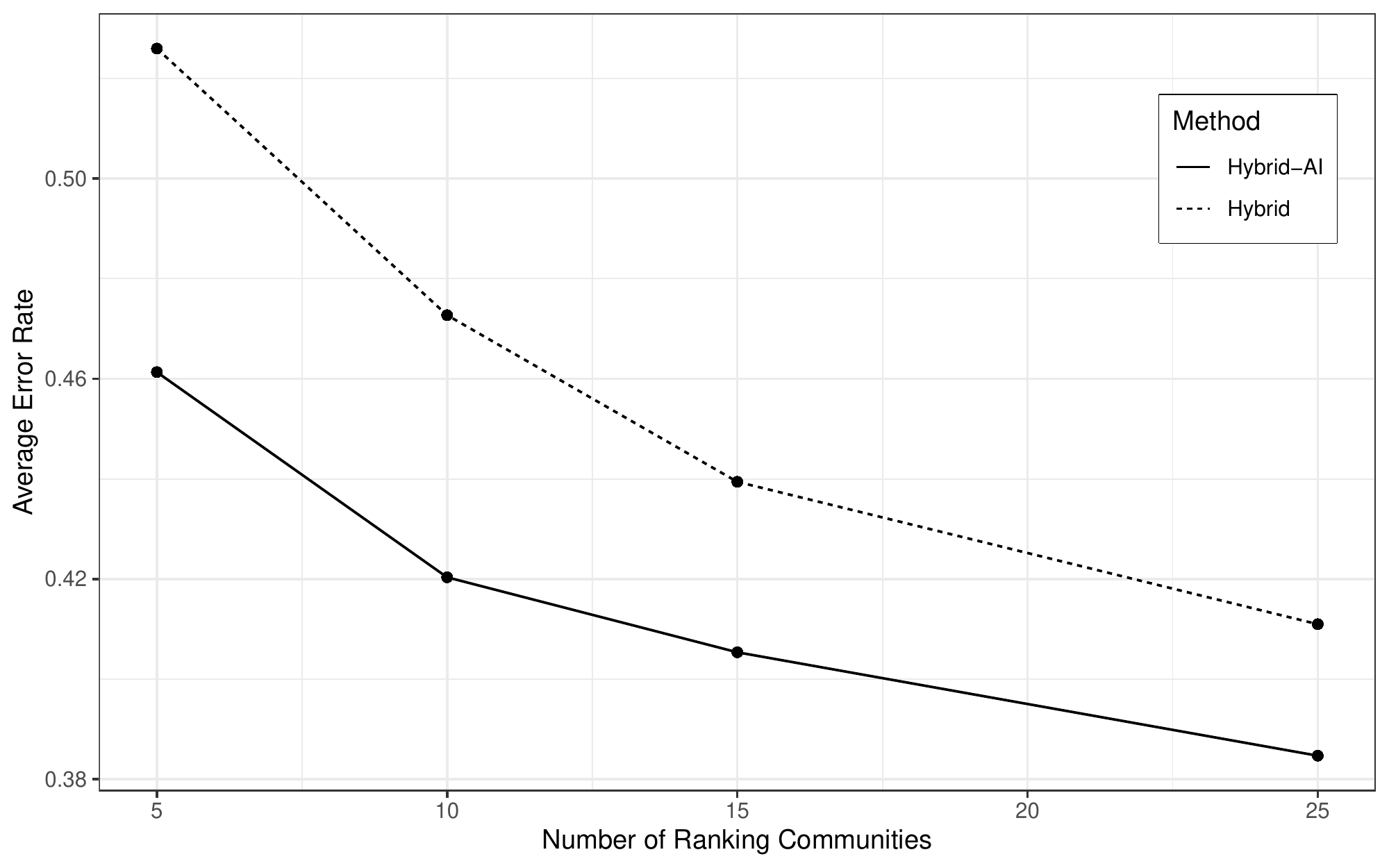}  
  \caption{Burkina Faso}
\end{subfigure}
\begin{subfigure}{.5\textwidth}
  \centering
  \includegraphics[scale=0.4]{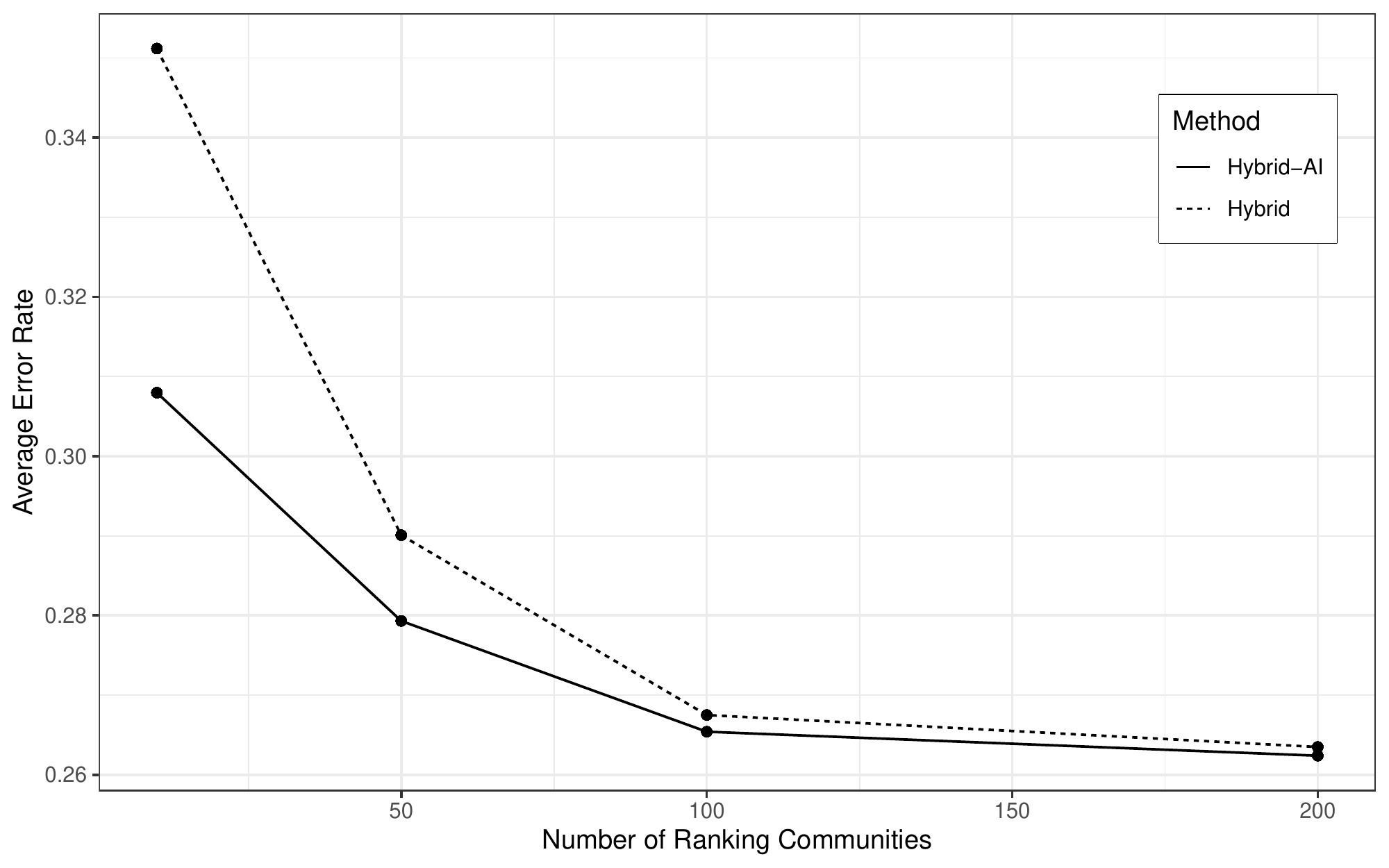}  
  \caption{Indonesia}
\end{subfigure}
\caption{Out-of-sample predictive performance of hybrid model with
auxiliary information}
\label{fig:er_AI}
\end{figure}

Figure \ref{fig:er_AI} presents results from the inclusion of auxiliary
information into our hybrid model (see Section \ref{ssec: aux_info} for 
discussion). 
Panel (a) presents the results from the Burkinab\'{e} data where the 
auxiliary component of the model is estimated using data from the 2007
Nouna Household Survey (654 households).
Panel (b) shows the results using the Indonesian data where the auxiliary
component of the model is estimated using data from those households 
randomly assigned to the PMT treatment (1,629 households).
For both countries, we then examine the out-of-sample predictive performance of 
this model (``Hybrid-AI'') relative to our basic hybrid model (``Hybrid'') for 
alternative training sample sizes.%
\footnote{Note that the sample size for the auxiliary model remains constant 
across all experiments.
We once again use 30 replications for each experiment.}
We find that including auxiliary information leads to non-negligible improvements 
in error rates for both countries, particularly for the smallest training sample 
sizes.%
\footnote{Recall that the auxiliary model informs $\delta$ indirectly through 
informing $\mu$ and $\Sigma$.
When the training sample size is small, the auxiliary model is relatively more
informative because the auxiliary sample size is fixed.
As the training sample size grows, the auxiliary model becomes relatively less
informative and the error rates from the model with auxiliary information 
eventually converge to those of the basic hybrid model.}
For example, for Burkina Faso, with five communities in the training sample, 
we see that the error rate is reduced by roughly seven percentage points.

\begin{figure}[!t]
  \centering
  \includegraphics[scale=0.4]{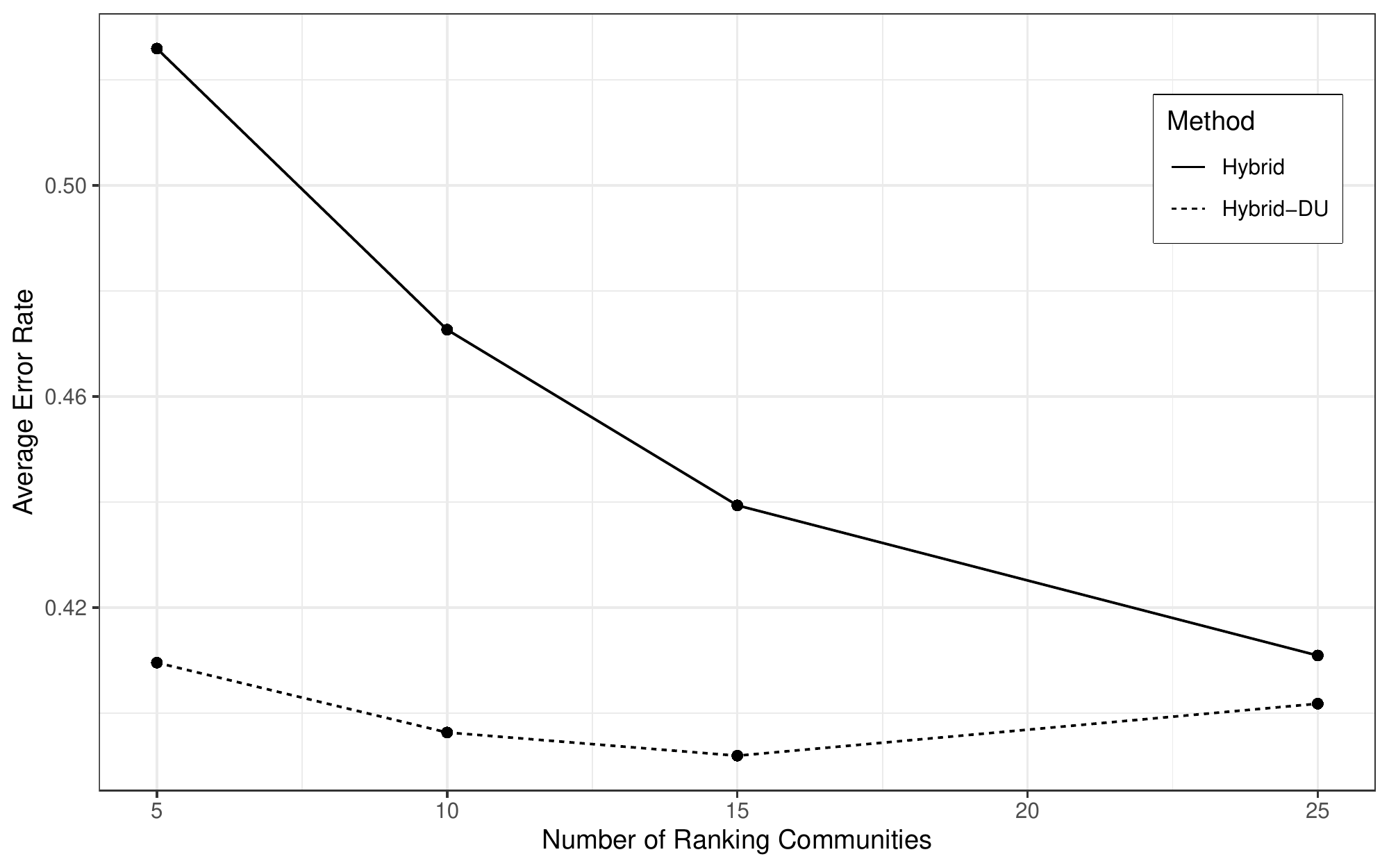}  
\caption{Out-of-sample predictive performance of hybrid model with
dynamic updating}
\label{fig:er_DU}
\end{figure}

Finally, Figure \ref{fig:er_DU} presents results associated with the dynamic
updating procedure discussed in Section \ref{ssec: dynamic_updating}.
We focus on the data from Burkina Faso because it is longitudinal.
Similar to the previous exercises, we examine the predictive performance of the
dynamically-updated model (``Hybrid-DU'') relative to the basic hybrid model 
(``Hybrid'') for alternative training sample sizes. 
In these simulation exercises, the basic hybrid model uses default priors
whereas the dynamically-updated model uses priors derived from the previous
period's posterior distribution.
Recalling that the Burkinab\'{e} training and testing samples correspond to the 
year 2009, the priors for the dynamically-updated model are thus derived from
the posterior estimated using 2007 data.
That is, we use the auxiliary sample (654 households) to inform the priors
used in the updating procedure.%
\footnote{Each run of the model uses the same priors estimated from the full 
auxiliary sample (i.e., the auxiliary sample size does not change as we vary
the training sample size).
Even though these experiments and the auxiliary information experiments both
draw on the auxiliary sample, they use the auxiliary sample in different ways.
Most notably, the auxiliary information experiments focus on household
expenditure data in the auxiliary sample whereas the dynamic updating experiments 
use the household ranking data.}

Figure \ref{fig:er_DU} shows that dynamic updating can lead to notable
improvements in error rates, especially when the training sample size is 
small.
For example, with five communities in the training sample, we find that 
dynamic updating reduces the average error rate by about 12 percentage points.
As the training sample size increases, the two models converge in terms of
targeting performance because the influence of the priors diminishes as the sample 
size grows.
Note that these results suggest that dynamic updating can be helpful for
economizing on data collection costs.
Specifically, the dynamically-updated model applied to the smallest sample size 
performs comparably to the basic model applied to the largest sample size.
Relative to ``memoryless'' updating procedures, dynamic updating thus allows 
social assistance programs to recertify targeting algorithms using a fraction of 
the sample size without sacrificing targeting performance.


\section{Conclusions}
\label{sec: conclusions}

PMT and CBT are two of the leading methods for targeting social assistance in 
developing countries.
While PMT relies only on verifiable information and imposes limited costs on 
potential beneficiaries, it is a highly centralized form of targeting that
neglects local information and preferences.
In contrast, CBT emphasizes local information and preferences by decentralizing
the targeting process, but as a consequence it can be costly to potential
beneficiaries and subject to elite capture.
We have thus proposed a hybrid targeting framework that incorporates many of the 
advantages of PMT and CBT while minimizing their main limitations. 
Our method resembles PMT in that beneficiary selection is based on a
weighted sum of verifiable sociodemographic features.
We nevertheless incorporate local information and preferences via weights that 
reflect the implied rate at which potential beneficiaries themselves substitute
sociodemographic features when making targeting decisions.

There are a few features of our method worth reiterating.
First, the method only requires community ranking data from a sample of eligible 
communities, thereby reducing the costs imposed on potential beneficiaries relative 
to CBT.
Second, we have extended our basic model to adjust for elite capture by explicitly
modeling the influence of elite connections on community rankings and purging 
these influences from the beneficiary selection process.
We have further extended the model to accommodate multiple rankers per community,
auxiliary information, and dynamic updating.
Finally, we have relied heavily on Bayesian methods, which facilitate the estimation 
of Thurstone-type models and enable many of our model extensions (e.g., dynamic 
updating).
Our Bayesian framework additionally provides a simple way to regularize the model's 
coefficients to improve out-of-sample predictions. 

We have illustrated our method using data from Burkina Faso and Indonesia.
The estimates from our hybrid model show that communities implicitly weight 
sociodemographic characteristics quite differently than PMT, with notable 
differences in sign and magnitude on features like roofing type, advanced education, 
and livestock ownership.
We also found that our method performs well in terms of predicting community
preference rankings out of sample.
Specifically, we found that our most basic model outperforms all benchmark models
and reaches error rates as low as 41 and 26 percent for the Burkinab\'{e} and 
Indonesian data, respectively.
These error rates are lower than what the standard PMT can achieve when 
predicting household expenditures out of sample.
Lastly, we demonstrated that further error rate reductions are possible when 
augmenting the basic model with auxiliary information or dynamically-updated 
priors, especially for smaller sample sizes.

Our results provide some guidance on the sample sizes necessary for applications
of our method.
The Indonesian data allowed us to consider a wider range of sample sizes and
we found that error rates stabilize at around 100 communities in that dataset
(see Figures \ref{fig:er_hybrid} and \ref{fig:er_AI}).
Given that the Indonesian data contains roughly nine households per community,
this suggests that a sample size of about 900 households is sufficient for the
initial implementation of the method.
If samples of this size are not feasible for the initial implementation (e.g., due 
to budget limitations), we recommend using the model with auxiliary information 
to improve targeting performance at smaller sample sizes (see Figure 
\ref{fig:er_AI}).
For repeated implementations, we recommend using our dynamic updating procedure,
which can help economize on data collection costs in subsequent rounds.
In particular, our results using the data from Burkina Faso suggest that 15 
communities or about 165 households is sufficient to recertify the 
targeting procedure (see Figure \ref{fig:er_DU}).%
\footnote{Recall that the data from Burkina Faso contains roughly 11 households 
per community.}

We conclude by discussing a few limitations of our method.
First, as is often the case with Bayesian methods, we use MCMC techniques
to sample from the posterior, which can be computationally demanding and
may require fine-tuning to achieve convergence.
Second, our method has some unique data needs, at a minimum requiring both (1)
data on community preference rankings for a sample of communities and (2) a census
capturing sociodemographic information for all eligible households.
It is nevertheless unclear whether our method is more or less costly than PMT or CBT, 
as a detailed understanding of relative costs will depend on the context and specifics 
of the implementation.%
\footnote{We are specifically referring to total costs, including administrative
costs, data collection costs, and the costs imposed on potential beneficiaries.
Recall that our method has one particular cost advantage, which is that it
will reduce the costs imposed on potential beneficiaries relative to CBT.}
Finally, we have relied on linear functional forms in all the models presented
in this paper, which likely restricts the predictive performance of our method.
In future work, we hope to extend the model to accommodate more flexible 
functional forms (e.g., regression trees).


\clearpage
\newpage
\bibliographystyle{apalike}
\bibliography{References}
\newpage


\newpage
\begin{appendices}
\section{}
\label{appendixa}

This appendix presents the computational details needed to implement our 
hybrid model. 
We first outline an MCMC procedure for sampling from the joint posterior
of our most general model specification (i.e., the model that accommodates
multiple rankers and auxiliary information).
This initial presentation uses default priors that can be used for any 
implementation of the method.
We then detail our dynamic updating procedure and show how the posterior
from any given implementation can be used to inform the priors for 
subsequent implementations.

\subsection*{MCMC specifics}

The joint posterior distribution for the most general form of our hybrid model 
can be written as follows:
\begin{align*}
    P(\tilde{z}, \alpha, \delta, & \; \omega, \gamma, \sigma^2_{\psi},\mu, 
    \Sigma \mid z, y) \\
    & \propto P(z \mid \tilde{z}) \; P(y \mid \gamma, \sigma^2_{\psi}) \; 
    P(\tilde{z} \mid \alpha, \delta, \omega) \; P(\alpha) \; 
    P(\gamma,\delta \mid \mu, \Sigma ) \; P(\mu) \; P(\Sigma) \; P(\omega) \;
    P(\sigma^2_{\psi}) \\
    &= \Bigg\{\prod_{k,r} \text{I}(\text{rank}(\tilde{z}_k^r) = z_k^r) \Bigg\}
    \Bigg\{\prod_{i,m}\text{N}(y_{im} \mid x_{im}\gamma, \sigma^2_{\psi})\Bigg\} \\
    & \times \Bigg\{ \prod_{r,i,k} \text{N}(\tilde{z}^r_{ik} \mid \alpha_{ik} +
    x_{ik}\delta, \omega^{-1}_r)\Bigg\} \Bigg\{\prod_{i,k} 
    \text{N}(\alpha_{ik} \mid 0, 1) \Bigg\} \\
    & \times \text{N}(\delta, \gamma \mid \mu, \Sigma) \; \text{N}(\mu \mid 0, 1) \;
    \text{Scale-inv-}\chi^2(\Sigma \mid 1,1)\\
    & \times \Bigg\{ \prod_{r}\text{Multinomial}(\omega_r \mid a_r) \Bigg\} 
    \text{Scale-inv-}\chi^2(\sigma^2_{\psi} \mid 1,1).
\end{align*}

\noindent
We cannot sample from the above joint posterior directly. 
We thus use an MCMC Gibbs sampler where we sequentially sample from each 
conditional posterior. 
For our applications, we run the algorithm for 4,000 iterations, but keep only 
the last $B=2,000$ iterations, discarding the initial burn-in samples. 
After initializing the parameters at reasonable starting values, we use 
Algorithm \ref{alg:mcmc}. 

\begin{algorithm}[h!]
\caption{Gibbs algorithm to sample approximately from $P(\tilde{z}, \alpha, 
\delta, \omega, \gamma, \sigma^2_{\psi},\mu,\Sigma \mid z, y)$}
\label{alg:mcmc}
\vspace{2mm}
For $b$ in 1 to B \textbf{do}:
\begin{enumerate}
  \setlength{\itemsep}{0pt}
  \setlength{\parskip}{0pt}
  \setlength{\parsep}{0pt}
    \item Draw $\tilde{z}^b \sim P(\tilde{z}\mid z, \alpha^{b-1}, \delta^{b-1},
    \omega^{b-1})$
    \item Draw $\alpha^b \sim P( \alpha \mid \tilde{z}^b, \delta^{b-1}, \omega^{b-1})$
    \item Draw $\delta^b \sim P(\delta \mid \tilde{z}^b, \alpha^b, \omega^{b-1},
    \mu^{b-1}, \Sigma^{b-1})$
    \item Draw $\gamma^b \sim P(\gamma \mid y, \sigma^{2, b-1}_{\psi}, \mu^{b-1},
    \Sigma^{b-1})$
    \item Draw $\omega^b \sim P(\omega \mid \tilde{z}^b, \alpha^b, \delta^b)$
    \item Draw $\sigma^{2, b}_{\psi} \sim P(\sigma^2_{\psi} \mid y, \gamma^b)$
    \item Draw $\mu^b \sim P(\mu \mid \delta^b, \gamma^b, \Sigma^{b-1})$
    \item Draw $\Sigma^b \sim P(\Sigma \mid \gamma^b, \delta^b, \mu^b)$
\end{enumerate}
\noindent end \textbf{do}. 
\vspace{2mm}
\end{algorithm}

For each of the steps in Algorithm \ref{alg:mcmc}, the specified conditional
posteriors are available in closed form. 
For several of the steps, it will be useful to represent the model specification 
in matrix form.
Let $X^K$ represent the matrix of sociodemographic characteristics that 
corresponds to the set of communities $K$ conducting community ranking 
exercises.
Further, let $X^M$ represent the matrix of sociodemographic features that
corresponds to the set of communities $M$ in the auxiliary component of the 
model.
We can then write:
\begin{align*}
    \tilde{z} &= X^R \alpha + X^K\delta + \eta \\
    y &= \phi + X^M\gamma + \psi 
\end{align*}
where $z^r_k = \text{rank}(\tilde{z}_k^r)$, $X^R$ is a random-effect design 
matrix, $\eta \sim \text{MVN}(0, \Omega)$, and $\psi \sim \text{N}(0, 
\sigma^2_{\psi}I)$.
Below we detail each of the conditional posteriors.

\textbf{Sampling $\tilde{z}$}: The conditional posterior distribution of 
$\tilde{z}$ can be written as:
\begin{align*}
P(\tilde{z} \mid \cdot) &\propto P(z \mid \tilde{z}) \; P(\tilde{z} \mid \alpha, 
\delta, \omega) \\
& \propto \Bigg\{ \prod_{k,r} \text{I}(\text{rank}(\tilde{z}_k^r) = z_k^r) \Bigg\} 
\Bigg\{ \prod_{r,i,k} \text{N}(\tilde{z}^r_{ik} \mid \alpha_{ik} + x_{ik}\delta, 
\omega^{-1}_r) \Bigg\}. 
\end{align*}

\noindent
The above implies a truncated normal conditional posterior for each element of 
$\tilde{z}$, where the bounds of the truncated normal are such that the original 
rank dictated by $z$ is preserved. 
Elements of $\tilde{z}^r_{ik}$ are drawn from their conditional posteriors 
sequentially, starting with the lowest ranked household within a community by 
ranker $r$ and ending with the highest ranked household. 

We represent the vector of ranks from ranker $r$ in community $k$ as $z^r_k = 
[z^r_{i_1 k}  \preceq z^r_{i_2 k} \preceq \hdots]$ where $i_1$ represents the index 
of the lowest-ranked household within community $k$, $i_2$ is the second lowest-ranked
household, and so on. 
We then draw for $h = 1,2,3, \hdots$ the following:
$$\tilde{z}^r_{i_h,k} \mid z, \tilde{z}^r_{-i_h,k}, \alpha, \delta \sim 
\text{N}(\alpha_{ik} + x_{ik}\delta, \omega_r^{-1}, \tilde{z}^r_{i_{h-1},k}, 
\tilde{z}^r_{i_{h+1},k})$$ 
where here $\text{N}(\cdot,\cdot,\cdot,\cdot)$ denotes a truncated normal distribution 
where the first two parameters represent the mean and variance, respectively, and the 
last two parameters represent the lower and upper bound, respectively. 

\textbf{Sampling $\alpha$}: The conditional posterior distribution of $\alpha$ can 
be written as:
\begin{align*}
P(\alpha \mid \tilde{z}, \delta, \omega) & \propto P(\tilde{z} \mid \alpha, \delta, 
\omega) \; P(\alpha ) \\
&= \Bigg\{ \prod_{r,i,k} \text{N}(\tilde{z}^r_{ik} \mid \alpha_{ik} + x_{ik}\delta,
\omega^{-1}_r)\Bigg\} \Bigg\{ \prod_{i,k}\text{N}(\alpha_{ik} \mid 0, 1) \Bigg\}.
\end{align*}
We can then draw from the conditional posterior as follows:
\begin{align*}
    \alpha \mid \tilde{z}, \delta, \omega &\sim \text{N}(\overline{m}_{\alpha},
    \overline{V}_{\alpha} )
\end{align*}
where
\begin{align*}
    \overline{V}_{\alpha}& = [(X^{R})^T\Omega^{-1}X^{R} + I]^{-1} 
\end{align*}
and
\begin{align*}
    \overline{m}_{\alpha} &= (\tilde{z} - X^K\delta)^T\Omega^{-1}X^R\overline{V}_{\alpha}
\end{align*}
represent the covariance matrix and mean of the multivariate normal distribution, respectively.

\textbf{Sampling $\delta$}: The conditional posterior distribution of $\delta$ can be 
written as:
\begin{align*}
    P(\delta \mid \cdot) &\propto P(\tilde{z} \mid \alpha, \delta, \omega) \; 
    P(\delta \mid \mu, \Sigma ) \\
    &= \Bigg\{ \prod_{r,i,k} \text{N}(\tilde{z}^r_{ik} \mid \alpha_{ik} + x_{ik}\delta,
    \omega^{-1}_r)\Bigg\} \text{N}(\delta \mid \mu, \Sigma).
\end{align*}
We can then draw from the conditional posterior as follows:
\begin{align*}
\delta \mid \mu, \Sigma, \tilde{z}, \omega &\sim \text{N}(\overline{m}_{\delta},
\overline{V}_{\delta})
\end{align*}
where
\begin{align*}
\overline{V}_{\delta} &= [(X^K)^T\Omega^{-1}X^K + \Sigma^{-1}]^{-1}
\end{align*}
and
\begin{align*}
\overline{m}_{\delta} &= [(\tilde{z} - X^R\alpha)^T\Omega^{-1}X^K +
\mu^T\Sigma^{-1}]\overline{V}_{\delta}
\end{align*}
represent the covariance matrix and mean of the multivariate normal distribution, respectively.

\textbf{Sampling $\gamma$}: The conditional posterior distribution of $\gamma$ can be 
written as:
\begin{align*}
    P(\gamma \mid \cdot) &\propto P(y \mid \gamma, \sigma^2_{\psi}) \; 
    P(\gamma \mid \mu, \Sigma ) \\
    &= \Bigg\{\prod_{i,m} \text{N}(y_{im} | x_{im}\gamma, \sigma^2_{\psi}) \Bigg\} 
    \text{N}(\gamma \mid \mu, \Sigma).
\end{align*}
We draw from the conditional posterior as follows:
\begin{align*}
\gamma \mid \mu, \Sigma, \tilde{z}, \omega &\sim \text{N}(\overline{m}_{\gamma},
\overline{V}_{\gamma})
\end{align*}
where
\begin{align*}
\overline{V}_{\gamma} &= [(X^M)^T(\frac{1}{\sigma^2_{\psi}}I)X^M + 
\Sigma^{-1}]^{-1}
\end{align*}
and 
\begin{align*}
\overline{m}_{\gamma} &= [y^T(\frac{1}{\sigma^2_{\psi}}I)X^M +
\mu^T\Sigma^{-1}]\overline{V}_{\gamma}
\end{align*}
represent the covariance matrix and mean of the normal distribution, respectively.

\textbf{Sampling $\omega$}: In the case of heterogeneous rankers, where we model 
a distinct $\omega_r$ for each ranker $r$, the conditional posterior probability 
that $\omega_r$ assumes the fixed value $w_l$ can be written as:
\begin{align*}
    P(\omega_r = w_l \mid \cdot) &\propto P(\omega_r)P(\tilde{z} \mid \alpha, \delta, 
\omega_r=w_l) \\
    &=a_{rl}\Bigg\{ \prod_{i,k}
    \text{N}(\tilde{z}^r_{ik} \mid \alpha_{ik} + x_{ik}\delta, w_l^{-1})\Bigg\}
\end{align*}
where $a_{rl}$ is the multinomial prior probability that ranker type $r$ has a 
quality weight equal to $w_l$ such that $l = 1, 2, 3$. 
Thus, the conditional posterior distribution of $\omega_r$ is
\begin{align*}
    \omega_{r} &\sim \text{Multinomial}(\overline{a}_{r1},\overline{a}_{r2},
    \overline{a}_{r3})
\end{align*}
where
\begin{align*}
    \overline{a}_{rl} &= \frac{a_{rl}  \underset{i,k}{\prod} \text{N}(\tilde{z}^r_{ik} 
    \mid \alpha_{ik} + x_{ik}\delta, w^{-1}_l) }{a_{r1}  \underset{i,k}{\prod}
    \text{N}(\tilde{z}^r_{ik} \mid \alpha_{ik} + x_{ik}\delta, w^{-1}_{1})  + 
    a_{r2}\underset{i,k}{\prod} \text{N}(\tilde{z}^r_{ik} \mid \alpha_{ik} + 
    x_{ik}\delta, w^{-1}_{2})  +  a_{r3}\underset{i,k}{\prod} 
    \text{N}(\tilde{z}^r_{ik} \mid \alpha_{ik} + x_{ik}\delta, w^{-1}_{3})}.
\end{align*}
In our applications, we set $(w_1, w_2, w_3) = (0.5, 1, 2)$ and 
$(a_{r1}, a_{r2}, a_{r3}) = (1/3, 1/3, 1/3)$ for each ranker $r$. 
In the next section, we provide an alternative specification based on the 
model with dynamic updating.

\textbf{Sampling $\sigma^2_{\psi}$}: The conditional posterior distribution of
$\sigma^2_{\psi}$ can be written as:
\begin{align*}
    P(\sigma^2_{\psi} \mid \cdot) &\propto P(y \mid \gamma, \sigma^2_{\psi}) \;
    P(\sigma^2_{\psi}) \\
    &= \left( \prod_{i,m} \text{N}(y_{im} \mid x_{im} \gamma, \sigma^2_{\psi})\right)
    \text{Scale-inv-}\chi^2(\sigma^2_{\psi} \mid 1,1).
\end{align*}
We draw from the conditional posterior as follows:
\begin{align*}
    \sigma^2_{\psi} \mid y, \gamma &\sim
    \text{Scale-inv-}\chi^2(\overline{n}_{\sigma^2_{\psi}},
    \overline{T}^2_{\sigma^2_{\psi}})
\end{align*}
where
\begin{align*}
    \overline{n}_{\sigma^2_{\psi}} &= 1 + \text{length}(y)
\end{align*}
and
\begin{align*}
    \overline{T}^2_{\sigma^2_{\psi}}&= \sum_{im}(y_{im} - x_{im}\gamma)^2 + 1
\end{align*}
are the degrees of freedom and scale parameters, respectively.

\textbf{Sampling $\mu$}: The conditional posterior distribution of $\mu$ can be 
written as:
\begin{align*}
    P(\mu \mid \cdot) & \propto P(\delta, \gamma \mid \mu, \Sigma) \; P(\mu) \\
    &= \text{N}(\delta \mid \mu, \Sigma) \; \text{N}(\gamma \mid \mu, \Sigma) \;
    \text{N}(\mu \mid 0, 1).
\end{align*}
We can then draw from the conditional posterior as follows:
\begin{align*}
    \mu \mid \cdot &\sim \text{N}(\overline{m}_{\mu}, \overline{V}_{\mu})
\end{align*}
where
\begin{align*}
    \overline{V}_{\mu}&= (2\Sigma^{-1} + I)^{-1}
\end{align*}
and
\begin{align*}
    \overline{m}_{\mu}&=[(\delta + \gamma)^T\Sigma^{-1}]\overline{V}_{\mu}
\end{align*}
denote the covariance matrix' and mean of the normal distribution, respectively.

\textbf{Sampling $\Sigma$}: In our formulation, $\Sigma$ is a diagonal covariance 
matrix where each diagonal element $\Sigma_{[p,p]}$ is the same. 
For simplicity, we refer to these diagonal elements as $\Sigma$. 
The conditional posterior can be written as:
\begin{align*}
    P(\Sigma \mid \cdot) &\propto P(\delta, \gamma \mid \mu, \Sigma) \; P(\Sigma) \\
    &= \text{N}(\delta \mid \mu, \Sigma) \; \text{N}(\gamma \mid \mu, \Sigma) \;
    \text{Scale-inv-}\chi^2(\Sigma \mid 1,1).
\end{align*}
We can then directly sample from the following:
\begin{align*}
    \Sigma \mid \gamma, \delta, \mu  &\sim 
    \text{Scale-inv-}\chi^2(\overline{n}_{\Sigma}, \overline{T}^2_{\Sigma})
\end{align*}
where
\begin{align*}
    \overline{n}_{\Sigma} &= 1 + \text{length}(\gamma) + \text{length}(\delta)
\end{align*}
and
\begin{align*}
    \overline{T}^2_{\Sigma}&= \sum_{p}(\gamma_p - \mu_p)^2 +
    \sum_{p}(\delta_p - \mu_p)^2 + 1
\end{align*}
represent the degrees of freedom and scale parameters, respectively.

\subsection*{Dynamic Updating}

We apply our dynamic updating procedure to the basic model outlined in Section 
\ref{sec: methods}.
Though the joint posterior $P(\theta \mid z_{1:t})$ of this model is unavailable in 
closed form, we have posterior samples $\{\theta^b: b = 1, \hdots, B\}$ that can 
be used to calculate the estimated posterior means $\hat{E}(\theta\mid z_{1:t})$,
posterior variances $\hat{V}(\theta\mid z_{1:t})$, and other summaries. 
In dynamic updating, our goal is to use these types of posterior summaries to 
approximate $P(\theta \mid z_{1:t})$, thus mathematically representing our beliefs 
after having seen some data. 
One simplification that will aid in the task of approximating a multivariate 
posterior distribution will be to assume independence. 
That is, we assume
\begin{align*}
    P(\delta, \omega) = P(\delta)P(\omega).
\end{align*}
The above implies independence between parameters with different roles. 
We further assume independence within elements of vectorized parameters  
(e.g., $P(\delta) = P(\delta_1)P(\delta_2)\cdots$).
In what follows, we present an approximation scheme for these parameters. 

\textbf{Approximating $P(\delta \mid z_{1:t})$}: Elements of $\delta$ are 
continuous and can be both positive and negative. 
\cite{gelman2013} notes that for parameters for which the posterior distribution 
is unimodal and symmetric, the normal distribution can be a convenient 
approximation. 
Thus, for the $p^{th}$ element of $\delta$, a normal distribution with mean
$\hat{E}(\delta_p \mid z_{1:t})$ and variance $\hat{V}(\delta_p \mid z_{1:t})$ would
provide a reasonable approximation to the true $P( \delta_p \mid z_{1:t})$. 
That is, we may set
$$\hat{P}(\delta_p \mid z_{1:t}) = N(\underline{m}_p, \underline{v}_p)$$
where
\begin{align*}
    \underline{m}_p &= \frac{1}{B}\sum_{b=1}^B\delta_p^b
\end{align*}
and
\begin{align*}
    \underline{v}_p &= \frac{1}{B}\sum_{b=1}^B(\delta_p^b-\underline{m}_p)^2.
\end{align*}

\noindent
A less informative prior would maintain the same central moment $\hat{E}(\delta 
\mid z_{1:t})$ as the mean of the normal prior, but inflate the variance to 
something larger than $\hat{V}(\delta \mid z_{1:t})$ by multiplying by some 
constant greater than one.

\textbf{Approximating $P(\omega_r \mid z_{1:t})$}: The parameter $\omega_r$ is 
discrete with a multinomial distribution. 
The hyperparameters $a_{r1}$, $a_{r2}$, and $a_{r3}$ of the multinomial distribution 
represent the relative probabilities with which $\omega_r$ takes on the possible 
values 0.5, 1, and 2. 
A reasonable approximation of $P(\omega_r \mid z_{1:t})$, then, might be
$$\hat{P}(\omega_r \mid z_{1:t}) =
\text{Multinomial}(\underline{a}_1,\underline{a}_2,\underline{a}_3),$$
where
\begin{align*}
\underline{a}_1& = \frac{1}{B}\sum_{b=1}^B \text{I}(\omega_r^b = 0.5),\\
\underline{a}_2& = \frac{1}{B}\sum_{b=1}^B \text{I}(\omega_r^b = 1),\\
\underline{a}_3& = 1-\underline{a}_1-\underline{a}_2.
\end{align*}
That is, we use the posterior samples to approximate the probability $\omega_r$ 
takes on various values.
A less informative prior would pull each of the probabilities $\underline{a}_1$,
$\underline{a}_2$, and $\underline{a}_3$ towards 1/3. 
Note that this is an especially important adjustment when \emph{a posteriori} 
$\omega_r$ takes on a particular value with probability zero. 
For discretely-valued parameters, a prior mass of zero forces a posterior mass of 
zero, which may be more restrictive than necessary.

\newpage
\section{}
\label{appendixb}

In this appendix, we detail our approach for comparing coefficients across models
by calculating the average marginal rate of substitution (MRS) for each 
sociodemographic variable. 
For a linear model, the MRS for variable $x_v$ with respect to variable $x_p$ can be 
written as
\begin{equation*}
\left| \frac{\Delta x_v}{\Delta x_p} \right| = \left| \frac{\delta_p}{\delta_v} \right| 
\end{equation*}
where $\delta_p$ and $\delta_v$ represent the coefficients that correspond to 
$x_p$ and $x_v$, respectively.
Rather than calculating the MRS for each pairwise comparison of covariates, we summarize 
the MRS for $x_p$ by calculating the \emph{harmonic mean} across all pairwise comparisons:
\begin{equation*}
\frac{\text{dim}(x)}{\sum_v \left| \frac{\Delta x_p}{\Delta x_v} \right|}
= \frac{\text{dim}(x)}{\sum_v \left| \frac{\delta_v}{\delta_p} \right|}
  \end{equation*}
where $\text{dim}(x)$ represents the number of sociodemographic variables in the 
model.

Simplifying the right-hand side of the above equation, we then have
\begin{equation*}
\frac{\text{dim}(x)}{\sum_v \left| \frac{\delta_v}{\delta_p} \right|}
= \frac{\text{dim}(x)}{\left| \frac{1}{\delta_p} \right| \sum_v \left| \delta_v \right|}
= \frac{\left| \delta_p \right|}{\frac{1}{\text{dim}(x)}\sum_v \left| \delta_v \right|}
= \frac{\left| \delta_p \right|}{\bar{\delta}}
\end{equation*}
where $\bar{\delta}$ represents the arithmetic mean of the absolute value of all 
coefficients in the model.
Combining the previous equations, we can then write
\begin{equation*}
\frac{\text{dim}(x)}{\sum_v \left| \frac{\Delta x_p}{\Delta x_v} \right|} 
= \frac{\left|\delta_p \right|}{\bar{\delta}}
\end{equation*}
meaning that the (harmonic) mean marginal rate of substitution for a given variable
$x_p$ can be calculated by dividing the absolute value of the associated coefficient 
by the arithmetic mean of all coefficients.
In terms of interpretation, the above tells us how many units, on average,
each variable must change to compensate for a one-unit increase in $x_p$
while holding the outcome variable constant.

Note that we summarize the MRS using the harmonic mean because the resulting
calculations only require that we divide each coefficient by the arithmetic mean of 
all coefficients, which is computationally stable.
An alternative approach would be to summarize the MRS using the arithmetic mean,
which would require that each coefficient be divided by the harmonic mean of all
coefficients.
This alternative approach is computationally unstable in that it would require summing
over the reciprocal of all coefficients, which is problematic if any coefficient is
near zero.
While we could avoid averaging altogether by choosing one coefficient as the 
numeraire, the choice of coefficient would be arbitrary and we would not be able
to interpret the MRS of the associated variable.

It is important to highlight a few additional details related to our calculations.
First, recall that Bayesian methods provide a posterior distribution for all 
coefficients.
As mentioned in Section \ref{ssec: basic_model}, we summarize this distribution by 
using the posterior mean as our estimated coefficients.
Second, we would like an understanding of the sign on each coefficient, so we 
simply ignore the absolute value function in the numerator and calculate
$\delta_p/\bar{\delta}$.
Finally, our calculations require that all variables are placed on a similar scale.
To accomplish this, we follow \citet{gelman2008scaling} and divide all
continuous variables by two times their standard deviation.
This follows from the observation that any binary variable with equal 
probabilities has a standard deviation of 0.5, meaning that a one-unit
increase in any binary variable roughly corresponds to an increase of two
standard deviations.
The standardization of the continuous variables thus ensures that all 
variables are on roughly the same scale.

\end{appendices}
\end{document}